\UseRawInputEncoding

\documentclass[preprint,1p,times,authoryear]{elsarticle}
\usepackage[pdftex]{color}
\usepackage[font=footnotesize,labelfont=bf]{caption}
\usepackage[font=footnotesize,labelfont=bf]{subcaption}
\usepackage{amsmath,amssymb,amsfonts}
\usepackage{algorithmic}
\usepackage{graphicx}
\usepackage{booktabs}
\usepackage{textcomp}
\usepackage{xcolor, soul}
\sethlcolor{lightgray}
\usepackage{comment}
\def\BibTeX{{\rm B\kern-.05em{\sc i\kern-.025em b}\kern-.08em
    T\kern-.1667em\lower.7ex\hbox{E}\kern-.125emX}}
    
\PassOptionsToPackage{hyphens}{url}\usepackage{hyperref}    
\usepackage{color}
\newcommand{\todo}[1]{}
\renewcommand{\todo}[1]{{\color{red} TODO: {#1}}}
\usepackage{lineno}

\AtBeginDocument{%
  \providecommand\BibTeX{{%
    \normalfont B\kern-0.5em{\scshape i\kern-0.25em b}\kern-0.8em\TeX}}}

\journal{Journal of Systems and Software}

\begin{document}

\begin{frontmatter}


\title{The Role of the Scrum Master in an Industry based University Course}


\author[Monash]{Kashumi Madampe\corref{cor}}

\author[UoA]{Zainab Masood}

\author[Monash]{Rashina Hoda}

\address[Monash]{Department of Software Systems and Cybersecurity, Monash University, Australia.}

\address[UoA]{Department of Electrical, Computer, and Software Engineering, The University of Auckland, New Zealand.}

\cortext[cor]{Corresponding Author}

\begin{abstract}
\textbf{Background:} Scrum is the most commonly used agile software development method, and the role of the Scrum Master (SM) in a Scrum environment is vital. Therefore, through an industry based university course for final year undergraduate and masters students, we aimed to give students both theoretical and practical understanding of the role of SM via hands-on experience on playing the role in real-world Scrum contexts. 
\textbf{Method: }We asked them (92 students) to share their experiences and learnings on the role of SM through reflective surveys and essays. Students participated in reflective surveys (311 survey responses) over 5 weeks in the course, and they submitted essays (92 essays) at the end of the course. We used a mixed-methods approach using Socio-Technical Grounded Theory analysis techniques and trend and regression based statistical analysis to analyse the survey responses and the essays.
\textbf{Findings: }We identified the key responsibilities and duties of the SM, common challenges faced by the SM and the team due to the role of the SM, root causes of the challenges, strategies used by the SM and the team to overcome the challenges, and the overall experience of the students. Based on the results, we present recommendations for educators.
\end{abstract}



\end{frontmatter}

\section{Introduction}
Agile software development is taught in university courses using different approaches comprising lectures, tutorials, and lab work with varying levels of practical experience.  While some have incorporated the use of interactive sessions such as board games \citep{Heikkila2016TeachingGame}, LEGO\textsuperscript{TM} \citep{Kropp2014TeachingCollaboration}, and game-based learning \citep{hoda2019using} to enhance the student learning experience, others have included project-based software development enabling students to gain experiential learning and develop software in quasi-real world project environments \citep{Devedzic2011TeachingStudy, Melnik2003IntroducingLearned, Paasivaara2015LearningTeams, Shukla2002AdaptingCourse}, and some delivered exclusively project-based courses without lectures \citep{Paasivaara2013TeachingScrum}. Agile courses are also offered with practical industry based project components \citep{Masood2018AdaptingContexts, Rico2009UseEducation}.

Scrum is the most popular agile method in industrial settings \citep{202014thAgile}. As the Scrum Master (SM) is integral to the Scrum method \citep{Deemer2012ThePrimer,2015TheGuide}, several studies have focused on the role of the SM  \citep{Bass2014ScrumProjects,Bolloju2018ProsStudy,Cowan2011WhenRunning,Davidson2016WhyMaster,Matturro2015SoftMasters,Noll2017ARole,Waugh2018IncreasingCertification,Yi2011ManagerMaster, yogi2020scrummaster} in industrial settings. However, research on the SM role in university contexts is limited \citep{Bolloju2018ProsStudy, Hans2017WorkProjects}. Even though how the students experience and perceive the role of the SM is not presented widely to both academic and industry audiences, we believe that it is important for both educators and industry collaborators to be aware of students’ reflections on the SM’s role that help to design courses better and ultimately prepare the students to better play the role when they enter the industrial workforce. Therefore, through this paper, we present students’ experiences (92 students) on the role of the SM based on the hands-on experience they gained through industry collaborated projects in a software engineering university course. 

The findings presented in this paper are derived from the analysis of data collected using reflective surveys (311 responses) over 5 weeks of the course and essays (92 essays). We used a mixed-methods approach with socio-technical grounded theory for data analysis \citep{Hoda2021Socio-TechnicalEngineeringb} techniques for the qualitative data and trend and regression based statistical analysis for the quantitative data. The contributions of this paper include, presenting the following, from the students' perspective:

\begin{itemize}
    \item Responsibilities and duties of the Scrum Master,
    \item Challenges faced by the Scrum Master and Team due to the role of the Scrum Master,
    \item Underlying causes of those challenges,
    \item Strategies to overcome the identified challenges, and 
    \item Our recommendations for educators to improve the learning experience of students
\end{itemize}


The rest of this paper is organised as follows. Section \ref{RW} presents an overview of the related work. The course design and setup is presented in Section \ref{TA}. The research design and approach comprising research questions, data collection, and analysis are elaborated in Section \ref{RD}. The findings are presented and discussed in Section \ref{Res}. Students' overall experience in the course and recommendation for educators are summarised in sections \ref{Exp} and \ref{Rcmd} respectively.

\section{Related Work}
\label{RW}


\subsection{Teaching Agile in University}
        Research on teaching agile in higher education has identified gaps in the agile curriculum, whereby start-ups struggle to find software engineering graduates well-trained and experienced in agile methods \citep{Devadiga2017SoftwareIndustry}. They argue for industry-focused curricula to be setup to align with the attributes of start-ups, such as innovation and cutting-edge software development. To address such industrial needs, educators teach agile methods in dedicated courses in universities across the world. Several studies have focused on the educators’ experiences of teaching agile in undergraduate and graduate courses, including studies on agile method specific experiences \citep{Bunse2004AgileEducation, Germain2005Engineering-basedStudy, Shukla2002AdaptingCourse}, practice-based experiences \citep {Kropp2014TeachingCollaboration,Paasivaara2015LearningTeams,Paasivaara2013TeachingScrum}, and overall experience \citep{Devedzic2011TeachingStudy, Rico2009UseEducation} including how agile practices are adapted to work in university contexts \citep{Masood2018AdaptingContexts}. Studies have defined key principles of teaching agile practices and recommended developing essential agile values such as communication and collaboration skills \citep{Kropp2014TeachingCollaboration}.
        
        
         However, teaching agile methods comes with a unique set of challenges. Extreme Programming (XP) is seen as easy to teach and learn but challenging to perform in agile projects due to limited student experience in traditional software development and limited guidance throughout the project \citep{Bunse2004AgileEducation}. Other studies have identified the challenges in applying global software engineering skills faced by students while teaching Scrum, strategies to overcome these challenges and the effectiveness of these strategies \citep{Paasivaara2013TeachingScrum}.
        
        At the same time, educators have provided different recommendations for teaching agile. Shukla and Willams suggest teaching both agile and traditional practices so that students would still learn techniques such as inspections, the Unified Modeling Language, and use cases \citep{Shukla2002AdaptingCourse}. Others suggest educators teaching specific agile practices such as pair programming, which is a part of the wider XP method, rather than attempting to teach the full XP method in undergraduate courses, arguing XP is an approach for talented and experienced teams \citep{Germain2005Engineering-basedStudy}.
        
        Multiple studies have provided recommendations to teach and apply agile methods in university courses effectively \citep{Devedzic2011TeachingStudy,Masood2018AdaptingContexts,Rico2009UseEducation}. Devedzic and Milenkovic shared the lessons learned over eight years of teaching agile \citep{Devedzic2011TeachingStudy}. They emphasise the positive role teachers play as a guide and to enforce skill and knowledge building. Small self-organised teams, short iterations, assigning the Scrum Master role, use of practical activities for learning, and acknowledging students varying capabilities were reported as some of the recommendations to effective agile learning and adoption.
    
    
\subsection{Practising Agile in University}
    
    Masood et al.’s research findings on students' practising agile methods in university contexts shows that students tailor standard agile practices to fit into the University settings, such as less frequent daily stand-ups, combined and sequenced sprint meetings, and employing a rotating Scrum Master \citep{Masood2018AdaptingContexts}. They recommend educators to organise for experienced tutors or Master students to play the SM role in the initial sprints, and elicit industry partners who are available and closely located to the University. Providing students dedicated places for their sessions, allocating time for daily stand-ups during classes and lab sessions, and a training period on practising agile techniques before project initiation are some of the other recommendations to help teach and practice agile. Supporting this, Rico and Sayani suggested an emphasis on technical proficiency, course work on agile methods, virtual collaboration, support through coaching and mentoring, involving knowledgeable customers, feedback, and teamwork for successful agile teaching in software engineering education \citep{Rico2009UseEducation}.
    
\subsection{Studies on the Scrum Master Role -- in Industry}
    
    Multiple studies have explored aspects of the Scrum Master (SM) role in industrial settings. Bass identified and described six duties performed by a SM in a large-scale enterprise software development \citep{Bass2014ScrumProjects}: a) \textit{process anchor} nurturing the adherence of Scrum, b) \textit{stand-up facilitator} ensuring team's active participation at stand-ups, c) \textit{impediment remover} allowing the developers to progress with their work, d) \textit{sprint planner} supporting the user story triage and general sprint planning, e) \textit{facilitator} coordinating with other SMs in the team and, f) \textit{integration anchor} facilitating code base merges by development teams. 
    
    In a recent study, Shastri et al. \citep{yogi2020scrummaster} describe the SM's role as involving everyday practices of \textit{facilitating, mentoring, negotiating, process adapting, coordinating}, and \textit{protecting}. They reported varying levels of the SM's involvement in agile practices and a positive relationship between the presence of the SM and the frequency with which agile practices were performed. Cowan highlighted the importance of SM’s role and identified characteristics of the SM, such as \textit{patience, coaching}, and \textit{encouragement}, which can lead the team towards self-organisation \citep{Cowan2011WhenRunning}. Similarly, Matturro et al.'s findings report that the most valued soft skills required to play the SM’s role effectively are \textit{communication skills, interpersonal skills, planning skills, teamwork, commitment}, and \textit{responsibility} \citep{Matturro2015SoftMasters}. Noll et al. indicate conflicts and tensions between roles of the SM and the Project Manager (PM) and suggest that PMs are more suitable to play the role of the Product Owner rather than the SM when transitioning from traditional to agile project management \citep{Noll2017ARole}.
    
    Others advocate the importance of senior management roles in Scrum adoption. Cowan’s research suggests involving the upper management roles such as Vice President to play the SM role to promote Scrum adoption \citep{Cowan2011WhenRunning}. Similarly, Davidson and Klemme endorsed the idea and explained their rationales in their study on \textit{why a CEO should think like a Scrum Master} \citep{Davidson2016WhyMaster}. They mentioned several benefits, with the CEO being the SM: speeding up innovation in software development projects, empowering value creation and customer involvement, removing barriers during development such as interventions by external managers, and altering the team’s remuneration structure. 
    
\subsection{Studies on the Scrum Master Role -- in University}
    A limited number of studies have explored the role of Scrum Master in academia. While teaching a software engineering course, Bolloju et al. compared a rotating SM to a fixed SM and did not find any significant differences \citep{Bolloju2018ProsStudy}. With a rotating SM, the burden on management was reduced and everyone realised responsibility of the SM. On the other hand, it is difficult to adjust to the new SM and switching slows down project development speed, especially when an inexperienced team member plays the role. Also, teams tend to compare the previous SM with the current SM. Similar disadvantages were reported for the dedicated SM but in different proportions, but a dedicated SM was seen to enhance leadership skills. 

    In a semester wide project-based course, Hans explored the impact of the student SM on quality and delivery time of student’s projects. During the course, the student SM was responsible for facilitating the scrum process. The preliminary findings report that student SM plays a positive role in supporting the team to meet project outcomes, increases project’s quality, and delivers the project within the deadline \citep{Hans2017WorkProjects}.

    Similarly, in a one-semester SM training course where students were trained using agile courses and communities of practice, Passivaara found that students preferred having agile coaches more in their course to train them. She also found that, SM community of practice well-suited when students were self-organised, they shared practices, tools, and also supported peers while playing the role. She also found that, students who had non-technical backgrounds played the role better than students with technical backgrounds. She highlights that social skills are important, and female students performed the role better \citep{paasivaara2021teaching}.
    
    Studies on the role of the SM have been conducted both in industrial and academic settings. Previous studies have identified the responsibilities SM plays while performing the role in university courses and the ways in which the SM role is played, i.e., rotating and dedicated SM and their pros and cons. However, there are limited dedicated studies focusing on the different aspects of the SM role in university contexts, particularly on the challenges the SM faces while playing the role, those faced by the team on account of incorporating the SM role, and strategies they adapt to overcome them.

\section{Course Design}
\label{TA}

SoftEng761 \textit{Agile and Lean Software Development} course was launched in 2013 at the University of Auckland, New Zealand by the third author as part of the Software Engineering undergraduate degree program. The course provides theoretical knowledge on agile methods and practices along with hands-on experience in developing a software application in a quasi-real world project context. Learning outcomes include self-organising teamwork, project management skills, research and reflection, and industrial practice.

\begin{figure}[]
        \centering
        \includegraphics[width=\textwidth]{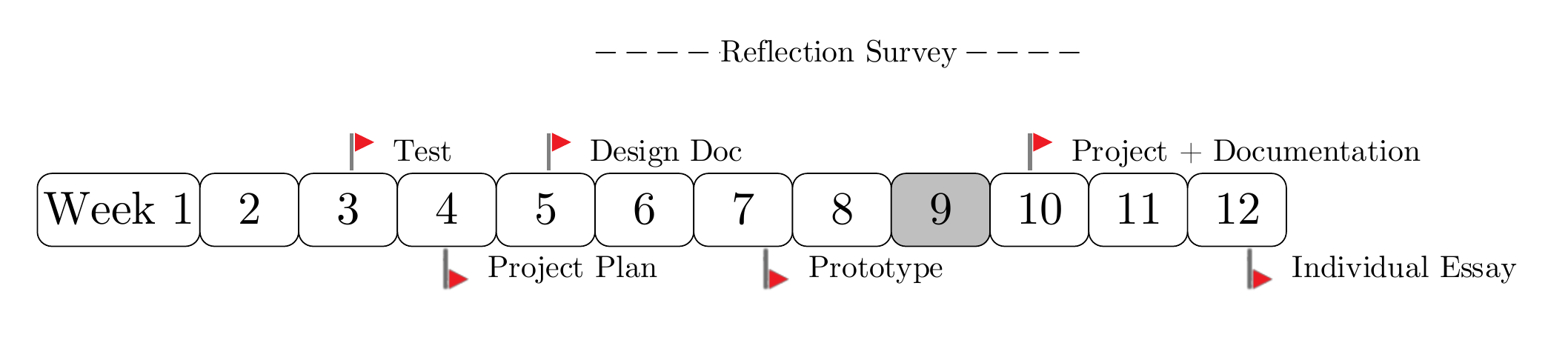}
        \caption{Course Timeline including \hl{Systems Week} (Systems Week: Final Year Project Exhibition Week)}
        \vspace{-0.5cm}
        \label{coursetimeline}
        \end{figure}

The course is taken up by students in the final year of Bachelor of Engineering (Honours), Master of Engineering Studies (pre-dominantly specialising in Software Engineering and some in Computer Systems Engineering), and Master of Information Technology. The course is structured around an average class size of 90 students with strong object-oriented programming skills, structured into 8 to 12 teams through a self-formation process for the project part of the course. The course workload is anticipated at 10 hours per week over 12 weeks. Over 417 students have taken the course in the last seven years, delivering 58 distinct projects to local industry partners. The course is structured into three sections, namely \textit{theory} in weeks 1-3, \textit{team project} in weeks 3-10, and \textit{individual research} in weeks 10-12. The course is assessed through a test on agile theory, interim deliverables of industry-based project (project plan, design documents, prototype, and final project code and documentation), and an individual reflective essay. Students filled reflection surveys throughout the duration of the project, as well as presented the project deliverables in class. Figure \ref{coursetimeline} shows the timeline of the course. Further information on the course design can be found in \citep{Masood2018AdaptingContexts}.

    \section{Research Design and Approach}
    \label{RD}
   A mixed method approach was employed to conduct the research. Primarily qualitative data and some supplementary quantitative data was collected through a series of reflection surveys weekly in weeks 6-10 (see Figure \ref{coursetimeline}) followed by an essay at the end of the course. The surveys and essay topics were designed to answer three research questions and seven sub-questions.\\

    \noindent \textbf{RQ1. What is the role of the Scrum Master in the university context?}
    \begin{description}
        \item 1a. What are the responsibilities and duties of the SM?
        \item 1b. What are the most commonly practised SM strategies?
        \item 1c. How effectively is the role of SM played?
    \end{description}

    \noindent \textbf{RQ2. What are the challenges faced in association with the SM role?}
    \begin{description}
        \item 2a. What are the challenges faced by the SM?
        \item 2b. What are the challenges for the team associated with the role of SM?
    \end{description}

    \noindent \textbf{RQ3. What strategies are used to overcome the challenges?}
    \begin{description}
        \item 3a. What are the strategies to overcome the challenges faced by SM?
        \item 3b. What are the strategies to overcome the challenges faced by the team associated with the role of SM?
    \end{description}

        \subsection{Data Collection}
        Data was collected through the course SoftEng761, comprising individual reflection surveys and essays. Reflection surveys were worth 5\% total, while the essay was worth 25\% of the course grades. Five reflection surveys were made available in weeks 6 to 10 of the course, resulting in n=311 total responses contributed by 92 students. The number of students participated in the surveys varied week by week. The essays written by each student as the last assessment item provided further insights into their experiences. Data collection was seamless as part of regular course assessment and did not burden the students with additional requests.
        
        Both qualitative and quantitative data was collected to answer the research sub-questions, as shown in Table \ref{tab:data-collection}. Qualitative data was collected through both reflection surveys and essays, whereas limited quantitative data was collected only through the reflection surveys.

        \begin{table}[ht]
        \resizebox{\textwidth}{!}{%
        \begin{tabular}{@{}ll@{}}
        \toprule
        \textbf{Research Sub-Questions}                                                                               & \textbf{Data} \\ \midrule
        1a. What are the responsibilities and duties of the SM?                     & \textbf{Qual}          \\
        1b. What are the most commonly practiced SM strategies?                    & Quan          \\
        1c. How effectively the role of SM is played?                               & Quan          \\ 
        2a. What are the challenges faced by the SM?                                                             & \textbf{Qual}          \\
        2b. What are the challenges for the team associated with the role of SM?                                 & \textbf{Qual}          \\ 
        3a. What are the strategies to overcome the challenges faced by SM?                                      & \textbf{Qual}          \\
        3b. What are the strategies to overcome the challenges faced by the team associated with the role of SM? & \textbf{Qual}          \\ \bottomrule
        \end{tabular}%
        }
        \caption{Data collection approaches to address the research sub-questions, primarily through qualitative (Qual) and supplemented by quantitative (Quan)}
        \label{tab:data-collection}
        \end{table}

        \paragraph{Reflection Surveys} The reflection surveys were designed and distributed on \textit{Qualitrics}\footnote{https://www.qualtrics.com/}.
        Some of the questions in the survey included,

        \begin{itemize}
        \item What is your team’s Scrum Master strategy? Options given: One dedicated SM, two dedicated SMs, rotating SM.
        \item Rate “The Scrum Master role was played effectively in the last sprint/week” on a scale 1-5 where 1 is strongly disagree, 2 is disagree, 3 is neutral, 4 is agree, and 5 is strongly agree.
        \item Based on your experience of the SM role in the last sprint/week, 
            \subitem -- what were the challenges for the team associated with this role?
            \subitem -- what were the challenges for the person playing this role (which may have been you)?
            \subitem -- what were the benefits for the team associated with this role?
            \subitem -- what were the benefits for the person playing this role (which may have been you)?

        \end{itemize}
        
        \paragraph{Essays} The essays focused on students’ perceptions of the challenges faced by the SM and strategies used to overcome those challenges through their project experience in the course. Students submitted an average of 4 -- 6 pages of written text, including examples of challenges and solutions.

    \subsection{Data Analysis}\label{subsec:DA}
\begin{figure}
            \centering
            \includegraphics[scale=0.6]{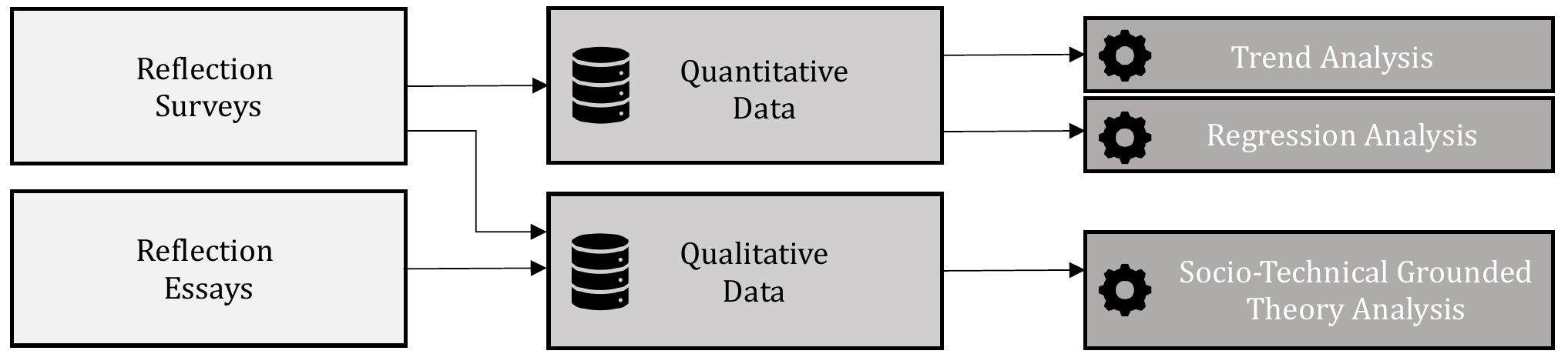}
            \caption{Data Analysis using Socio-Technical Grounded Theory for data analysis \citep{Hoda2021Socio-TechnicalEngineeringb} techniques for qualitative data and Trend and Regression Analysis for quantitative data}
            \label{fig:dataAnalysis}
        \end{figure}

        The primary qualitative data was analysed using socio-technical grounded theory \textit{(STGT) for data analysis} \citep{Hoda2021Socio-TechnicalEngineeringb} techniques such as open coding and constant comparison, while the supplementary quantitative data was analysed using trend and regression analysis (see Figure \ref{fig:dataAnalysis}). 
        
        \paragraph{Qualitative Data Analysis}
        \textit{STGT for data analysis} techniques of open coding and constant comparison were used to analyse the data collected from the open-ended survey questions and the text of the essays to uncover the duties and responsibilities of SM, challenges, strategies, underlying causes, and relationships among challenges and strategies. Qualitative data from reflection surveys in Qualtrics were open-coded. Essays were grouped into team level, open-coded, and recorded in Excel sheets. Open codes derived from the Qualtrics surveys were transferred to same Excel sheets by merging to respective teams and constantly compared to identify key patterns in the data.
        
        \paragraph{Quantitative Data Analysis}
        Quantitative data collected through reflection surveys were analysed using \textit{Microsoft Excel}. Week-wise analysis to find the trends of practised SM strategies (Section \ref{sec:strategies}) and \textit{effectiveness} of the SM role (Section \ref{sec:effectiveness}) and presented through plotting bar graphs. Similarly, taking data of all weeks into consideration, student perception on playing the role vs effectiveness of playing the role was analysed. Further analysis to identify the existence of relationships between SM Strategy and effectiveness, and student satisfaction with customers and with project outcomes was conducted via regression analysis.

\section{Findings}
\label{Res}
\subsection{Role of the Scrum Master in University Contexts (RQ1)}

        \subsubsection{Responsibilities and Duties of the Scrum Master (sub-RQ 1a)}
\begin{figure}[t]
        \centering
        \includegraphics[scale=0.3]{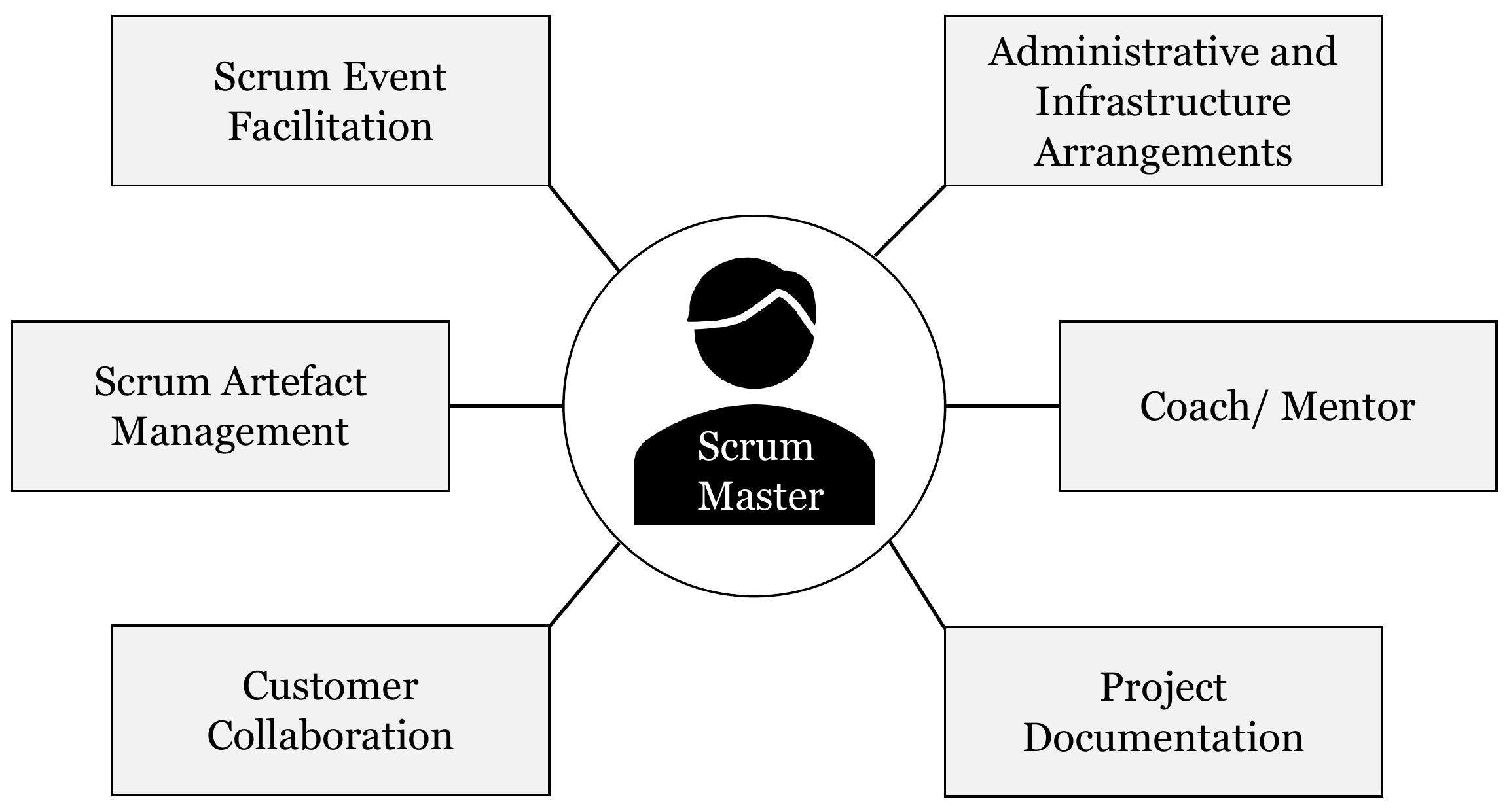}
        \caption{Responsibilities and Duties of the Scrum Master in University Context}
        \label{fig:responsibilities}
    \end{figure}

    During the course of SoftEng761, SM was reported to perform a number of responsibilities and duties, grouped into six key areas: (a) \textit{Scrum events facilitation}, (b) \textit{Scrum artefacts management}, (c) \textit{customer collaboration}, (d) \textit{administrative and infrastructure arrangements}, (e) \textit{coaching/mentoring}, and (f) \textit{project documentation}, depicted in Figure \ref{fig:responsibilities} and described below. Similar to industrial SM, many of these duties were related to facilitating application of Scrum practices and events throughout the SoftEng761 project. However, there were some duties which were more academic-oriented, such as arranging rooms for collective development team sessions and submitting the project deliverable to the teaching staff.
        
        \paragraph{\textbf{Scrum Events Facilitation}}
        In our industry project based course, the Scrum Master was mainly responsible for facilitating the Scrum ceremonies or events such as daily stand-ups, sprint reviews, and retrospectives. These responsibilities involved initiating the stand-ups, time-boxing them, and ensuring that everyone in the team gets their turn. Similarly, when running the retrospectives, SM would prompt team members to address the agenda, translate retrospective outcomes into actionable points for improvements. SM also played an important part in guiding the teams that had a combined Sprint review, planning, and retrospective session to keep track of the agenda, time and progress. It was also reported that these duties were predominantly limited to initial sprints when the teams were establishing team practices. The SM also helped the team in determining individual efforts (e.g. in story points, hours) for each item in the Sprint backlog and Sprint velocity.
        
         \begin{center}
           \textit{ ``They [SM] ensured that the Scrum ceremonies were formally conducted and that the team, including the product owner, maintained appropriate and   focused  conversations during   these   ceremonies.''} -- Team Member (Team 7)
        \end{center}
        
        \paragraph{\textbf{Scrum Artefacts Management}}
        The SM was solely responsible for organising the Scrum artefacts in most of the teams. This included setting up and maintaining Scrum artefacts such task boards, the product/sprint backlogs, updating the burn-down charts using different tools through GitHub wiki content, Trello boards. For some of the teams, it was the SM who facilitated self-assignment, monitored the tasks creation, and ensured that everyone in the team had a task assigned to them and helped track them to completion.
        \begin{center}
           \textit{ ``Maintenance of our story tracking tool [Trello] was also a task that  the  Scrum  Master  was  heavily  involved with.''} -- Team Member (Team 7)
        \end{center}

        \paragraph{\textbf{Administrative and Infrastructure Arrangements}} 
        The SM organised regular meeting times and locations for Sprint ceremonies and co-located development time slots. Other duties include room bookings through the teaching staff, coordinating with the customer for their availability, coordinating with the lecturer and teaching assistants regarding the project, submitting due deliverables to the teaching staff, setting up a Slack channel or a bot for the team’s communication.
        \begin{center}
          \textit{  ``The SM  organised  regular  meeting  times  and  locations  for  sprint ceremonies and co-located development.''} -- Team Member (Team 9)
        \end{center}
        
        \paragraph{\textbf{Coaching and Mentoring}} 
        It was SM’s responsibility to ensure that the team followed the Scrum process appropriately through duties such as removing obstacles that stood in the team’s way, evoking different ideas for the team to try out, sharing innovative ways for running better retrospectives. Similarly, they contributed in resolving team related issues, driving and facilitating team discussions, building mutual understanding within team and customer, handling disagreements within team, and providing constant reminders and words of encouragement to keep team members motivated. The SM was seen to not only guide the team, but also the Product Owner (PO).
        \begin{center}
           \textit{ ``The SM  guided  the  Product  Owner  and  the  team  on  how  to  use  the tools and coached  the  Product Owner  on  writing  Product Backlog items.''} -- Team Member
        \end{center}
        
        \paragraph{\textbf{Customer Collaboration}}
        The SM was the main point of contact for the customer represented by the PO and was responsible to facilitate the communication between the team and the customer for some teams. The duties performed by the SM in this case included collecting and passing on requirement related inquiries from the team to the PO, acting as a mediator when necessary.
        \begin{center}
           \textit{ ``Our   SM was   the official correspondent  of  the  team  for  contacting the  Product Owner. The tasks included collecting  inquiries  that  the  team  may  have  for the  requirements  of  the  application  and forwarding  them  to the Product Owner through emails.''} -- Team Member (Team 9)
        \end{center}
        
        Customer collaboration also included conveying the customer’s expectations to the team and assisting the customer in prioritising the Product Backlog items to maximise the business value. 
        
        
        \begin{figure}[t]
        \centering
        \includegraphics[scale=0.4]{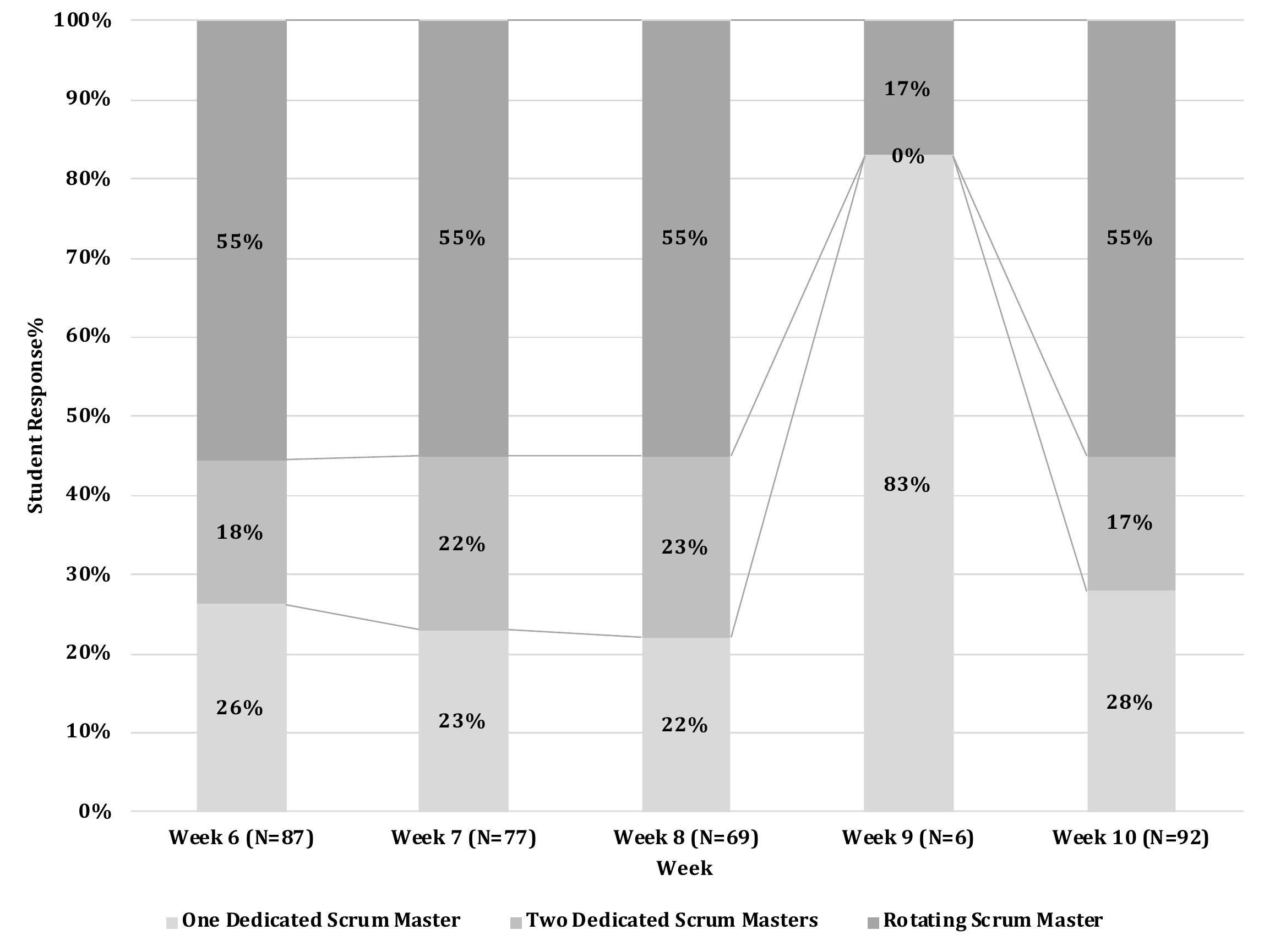}
        \caption{Use of Scrum Master strategies (one dedicated SM, two dedicated SMs, and rotating SM) through the project}
        \label{fig:strategies}
    \end{figure}
        
        \paragraph{\textbf{Project Documentation}}
        In some teams, the SM was also responsible for creating most of the project documentation as per the course’s requirements, while the other team members were involved in the development duties. Apart from project documentation, the SM was responsible for recording and circulating meeting notes, the PO’s feedback, and the team’s reflections from retrospectives. 
        \begin{center}
           \textit{ ``I [SM] was   solely responsible  for  creating  most  of  the project documentation,  other team  members  were  able  to  focus on  development.''} -- Scrum Master (Team 4)
        \end{center}
  
    \subsubsection{Most Commonly Practised Scrum Master Strategies (sub-RQ 1b)}
    \label{sec:strategies}
    Considering the SM strategy practised, as shown in Figure \ref{fig:strategies}, the \textit{rotating SM} strategy -- where individual members took turns to play the SM role -- was employed approximately 55\% from weeks 6 to 8 and in week 10. Some teams used \textit{two dedicated SMs} as a way to back up when one SM was unavailable. A drastic fall down to 17\% was seen in week 9 where the \textit{one dedicated SM} strategy peaked (otherwise below 30\% in other weeks). The strategy of \textit{two dedicated SMs} playing the role was below 24\% in most weeks and nil in week 9. Week 9 was a special case as it was \textit{Systems Week} (see Fig \ref{coursetimeline}) in the Faculty dedicated to an intensive, interdepartmental Engineering project, a major commitment outside the agile and other courses. Students were not expected to work actively on the agile project in this week and the responses for the reflection surveys dropped drastically (n=6) as the survey was also made optional for that week. Overall, teams preferred having a \textit{rotating SM strategy} the most, followed by a \textit{single dedicated SM} while \textit{two dedicated SMs} were used sparingly.

    \subsubsection{Effectiveness of Scrum Master's Role (sub-RQ 1c)}
    \label{sec:effectiveness}
\begin{figure}[]
        \centering
        \includegraphics[scale=0.4]{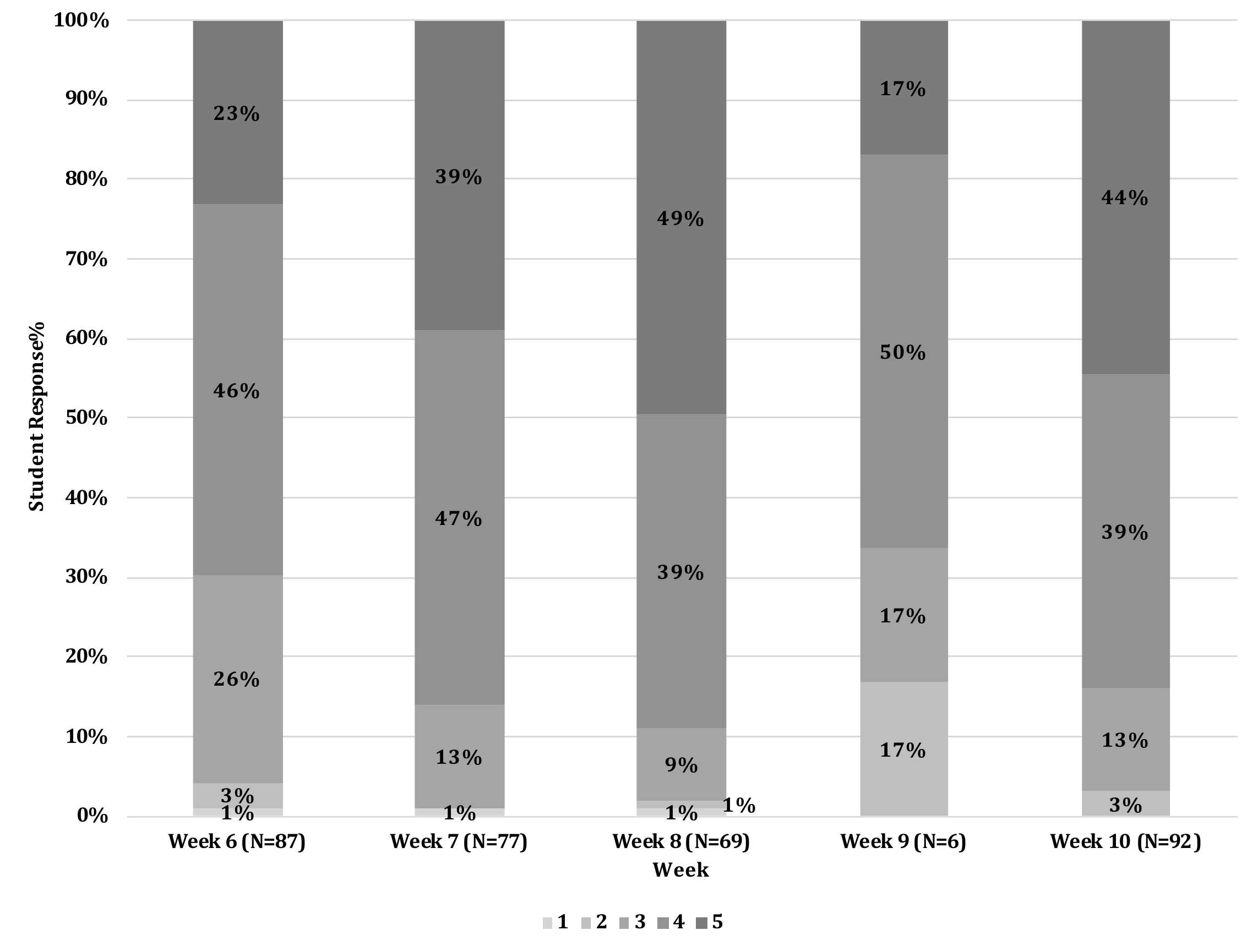}
        \caption{Effectiveness of Scrum Master's Role (1=lowest; 5=highest)}
        \label{fig:effectiveness}
    \end{figure}

    The responses given by the students on the effectiveness of the SM role is shown in Figure \ref{fig:effectiveness}. Rating was performed on a scale of 1-5 where 1 is the lowest and 5 is the highest. While the Systems Week (week 9) impacted the project work, overall the teams gained confidence in playing the SM role over the weeks, where 69\% reported SM effectiveness as high (rating of 4 and 5) in week 6 and 83\% percent rated the SM effectiveness as high by the end in week 10.

\subsection{Challenges faced in association with the SM role (RQ2)}


    \subsubsection{Most Common Challenges Faced by the Scrum Master (sub-RQ 2a)}
\begin{figure}[t]
        \centering
        \begin{minipage}{.5\textwidth}
        \centering
        \includegraphics[width=.85\linewidth]{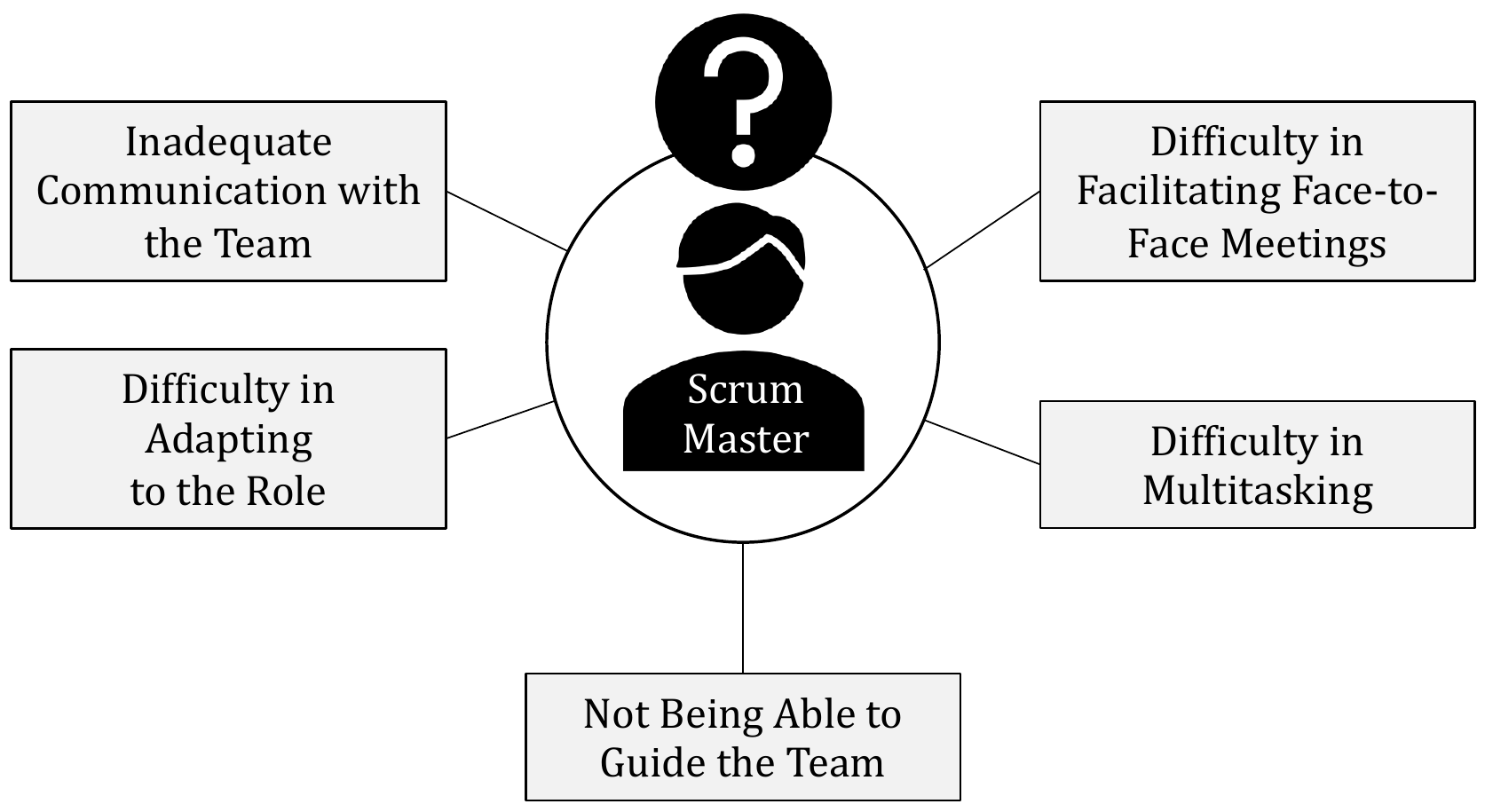}
        \caption{Challenges Faced by the Scrum Master}
        \label{fig:SMChallenges}
        \end{minipage}%
        \begin{minipage}{.5\textwidth}
        \centering
        \includegraphics[width=.9\linewidth]{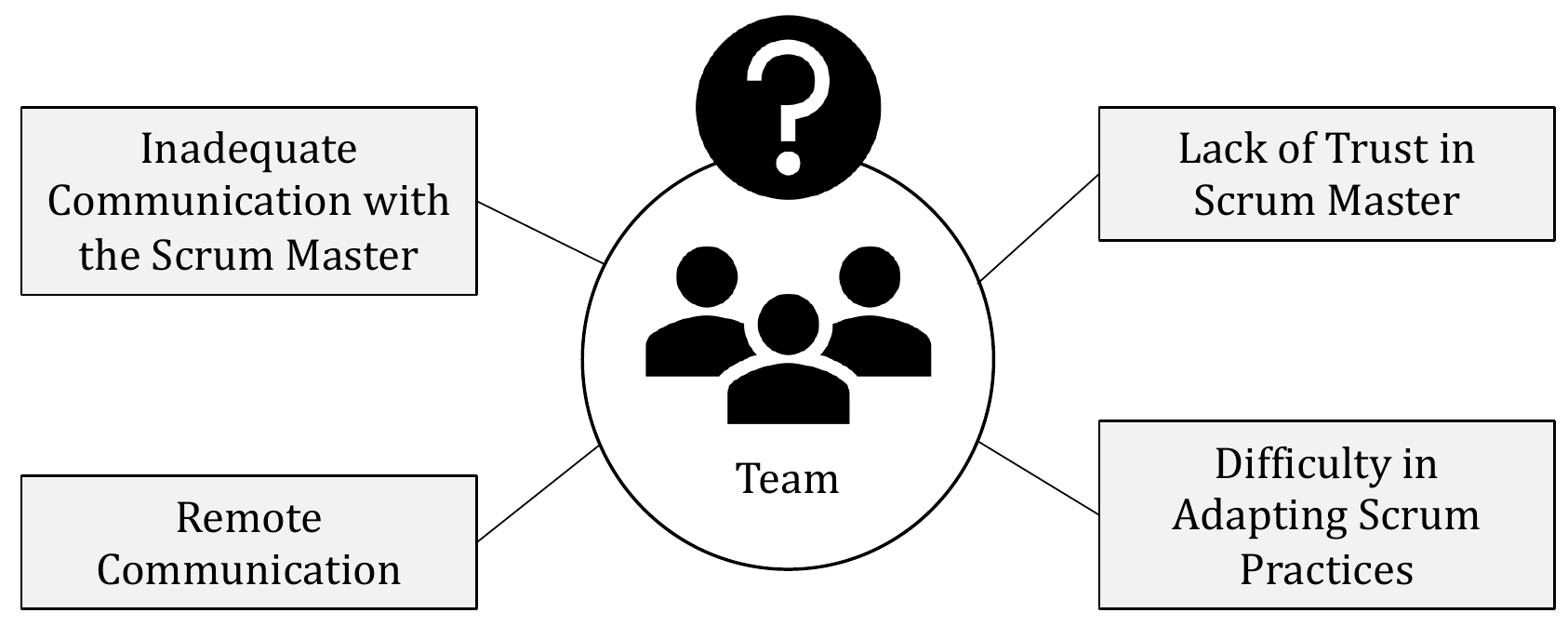}
        \caption{Challenges Faced by the Team}
        \label{fig:TeamChallenges}
        \end{minipage}
    \end{figure}

    Our analysis shows that the SM faces numerous challenges (depicted in Figure \ref{fig:SMChallenges}): inadequate communication with the team, difficulty in adapting to the role, not being able to guide the team, difficulty in facilitating face-to-face meetings, and difficulty in multitasking were the most common challenges faced by the SM reported by the students.

    \paragraph{\textbf{Inadequate Communication with the Team}}
     Communication was reported as a challenge for the SM since teams tended to carry on their discussions by themselves, purposefully or inadvertently ignoring the SM. This was particular evident in the online tools used in remote communication. Since the motivation of the teams was to deliver the software product, it can be assumed that the teams focused more on discussing development related work than process related topics often focused on by the SM. It can also point to the democratic nature of the team, where members were empowered to communicate directly and not feel the need to mediate through the SM.
     \begin{center}
       \textit{ ``Sometimes the voice of the scrum master was lost in the crowd. Especially on slack chats.''} -- Team Member (Team 9)
    \end{center}
    
    \paragraph{\textbf{Difficulty in Adapting to the SM Role}}
    When practising the \textit{rotating SM} strategy, different team members get the opportunity to play the role. Therefore, a transition period is inevitably needed in getting adapted to the role which is natural and acceptable. The effectiveness of the transition seemed to vary with  personal traits such as enthusiasm and professionalism.
    \begin{center}
       \textit{ ``It is challenging with the rotation of the scrum master because each person approaches the job with different levels of enthusiasm and professionalism.''} -- Team Member (Team 4)
    \end{center}
    
    \paragraph{\textbf{Not Being Able to Guide the Team}}
    The SM not being able to guide the team makes it difficult for the team, but also for the SM himself/herself. A number of issues can contribute to this, as discussed later.
    \begin{center}
       \textit{ ``...sometimes the Scrum Master does not know what to do in specific situations when the team becomes confused as to what process should follow to conform to sticking to agile practices...''} -- Team Member (Team 5)
    \end{center}
    
    Lack of experience and lack of authority of the SM on some occasions resulted in a chaotic environment as reported by some students. Two extreme were seen, one where the SM attempted to micromanage the team and consequently became disruptive to the team practising self-organisation. And the other, where the SM left the team alone entirely in a bid to empower them. These reasons made some students doubt the role of SM in university contexts.
    
    \begin{center}
        \textit{``...SM role to be quite unnecessary in a university context. This is because of the SM's lack of authority and inexperience...difficulty of choosing the line between micro managing and being disruptive to the self-organizing teams and not doing enough to empower the team and lead them onto the right track.''} -- Team Member (Team 7)
    \end{center}

    \paragraph{\textbf{Difficulty in Multitasking}}
    In a university context, some students playing the SM role found it difficult to multitask across the agile course and their other courses. They found the SM role particularly demanding and stressful as they were not only responsible for organising their own project schedule, multitasking across courses, but also reminding others of the various ceremonies and events.
    
    \begin{center}
        \textit{``[The] SM [role] is more stressful than I expected, especially because I had many other projects on and had to remember to update everyone on the meetings and doing their retro tasks. ''} -- Scrum Master (Team 11)
    \end{center}
    
    
    \subsubsection{Most Common Challenges Faced by the Team (sub-RQ 2b)}
    How the role of the SM was enacted in practice created some challenges for the team, including: inadequate communication with the SM, lack of trust in the SM, remote communication, and difficulty in adapting Scrum practices as depicted in the Figure \ref{fig:TeamChallenges}.
    
    \paragraph{\textbf{Inadequate Communication with the Scrum Master}}
    
    Inadequate communication was not only mentioned by as a challenge faced by the SM but also by the team. Some teams experienced lack of coordination amongst themselves as the SM did not communicate requirements effectively and some team members were left without tasks to perform in the absence of effective communication and coordination.
    
    \begin{center}
           \textit{ ``The main challenge for the team was that we were uncoordinated and often had times where we didn't know who was doing what, and some people did not have work to do at times.''} -- Team Member (Team 2)
        \end{center}
    
    \paragraph{\textbf{Lack of Trust in the Scrum Master}}
    Sometimes the SM instructed the team not to move into development, 
    and that causes members to be frustrated. Not explaining the rationales behind such recommendations can lead to a lack of trust in the Scrum Master. 
    
    \begin{center}
       \textit{ ``It was often frustrating when we wanted to get into development but were made to wait by the scrum master.''} -- Team Member (Team 8)
    \end{center}

    \paragraph{\textbf{Remote Communication}}
    Remote communication makes the team members lose their focus. Technical issues such as interrupted internet connections, background noise, and other distractions such as giving priority to other work whereas priority was required to be given to the meeting when having the meeting have challenged the team members. The selection of the place to hold the meeting also had distracted the teams as the environment was not set to have a proper meeting. Therefore, in such cases, SM is supposed to get the team back to an agile mindset but had not been met as reported by the students.
    \begin{center}
       \textit{ ``The meetings frequently got interrupted due to poor Wi-Fi connection. It was sometimes to block outside distractions, for example, noise in the leech [meeting]  area and a team member was distraction others with irrelevant course work from another course during a Sprint meeting.''} -- Team Member (Team 1)
    \end{center}
    
    
    \paragraph{\textbf{Difficulty in Adapting Scrum Practices}}
     One of the Scrum practices is having timeboxed events, such as daily stand-ups. It is the responsibility of the SM to adhere to these principles. Teams might find the time is not enough for them to fulfil the purpose of the event. This may make it difficult for the SM to play his/ her role as well. If time boxed meetings go beyond the time limit, it hints that the team might be having unnecessary discussions which were not supposed to be done during the particular event. For example, when the team tries to discuss the development activities in depth during a daily stand-up, it deviates from the objective of the meeting and also at the same time the team tends to think that they are not given enough time. 
        \begin{center}
           \textit{ ``Sticking to the Scrum meeting time boxes was probably the biggest challenge. The Scrum Master was trying to make sure we stuck to the time boxes of each meeting however, the group wanted to be thorough. These goals were a bit at odds with each other.''} -- Team Member (Team 6)
        \end{center}

\subsection{Strategies to Overcome the Challenges Faced in Association with the SM Role (RQ3)}
    
    \subsubsection{Most Common Strategies Used by Scrum Master (RQ3a)}
    To overcome the challenges faced by the SM, students have identified the strategies: having a dedicated SM, gaining Scrum knowledge, training SM, and having a dedicated location for meetings as shown in Figure \ref{fig:strategies}. These are based on the lessons they learnt through the project.

    \begin{figure}[t]
        \centering
        \begin{minipage}{.5\textwidth}
        \centering
        \includegraphics[width=.8\linewidth]{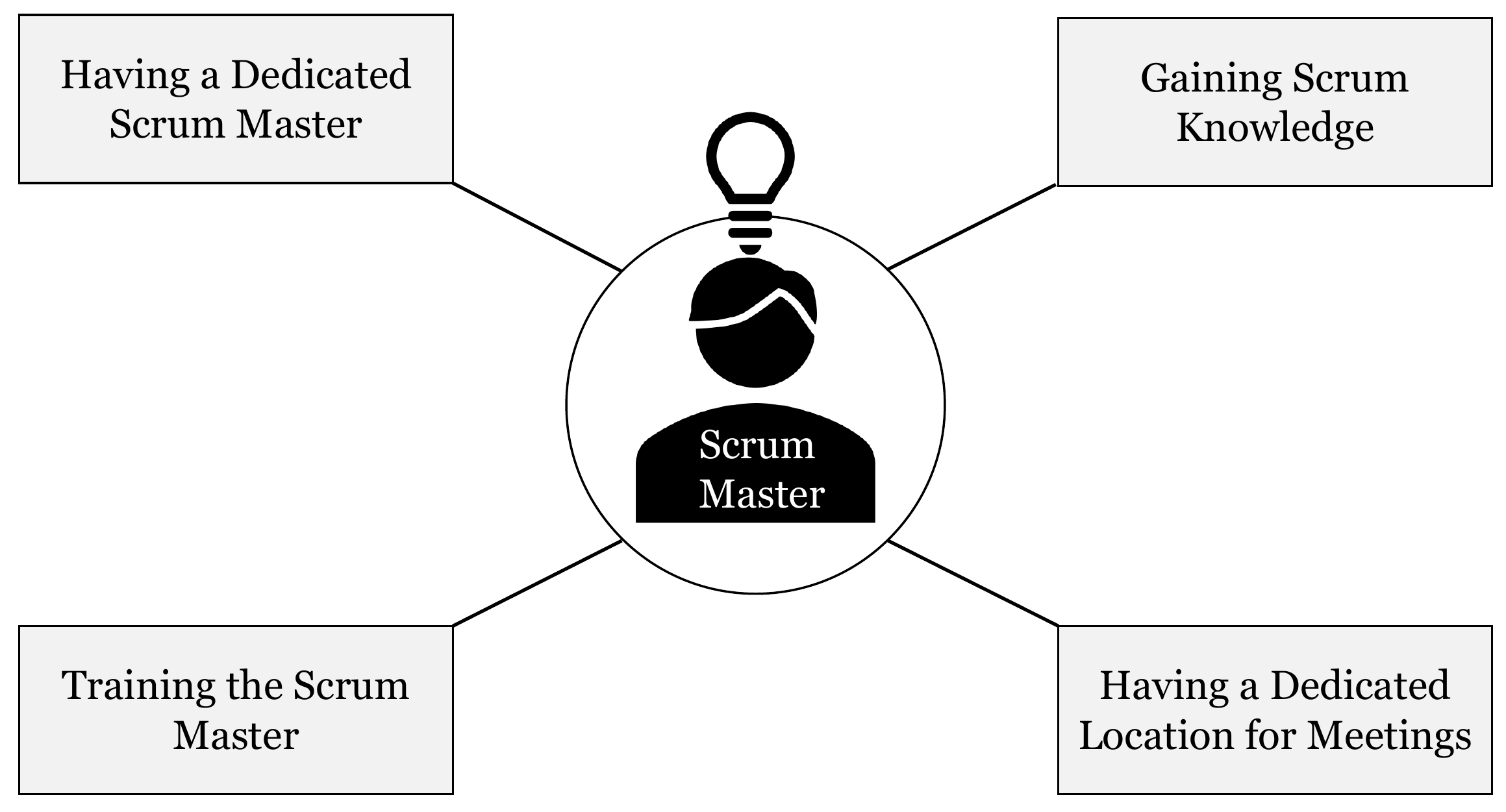}
        \caption{Strategies Used by Scrum Master}
        \label{fig:SMStrategies}
        \end{minipage}%
        \begin{minipage}{.5\textwidth}
        \centering
        \includegraphics[width=.9\linewidth]{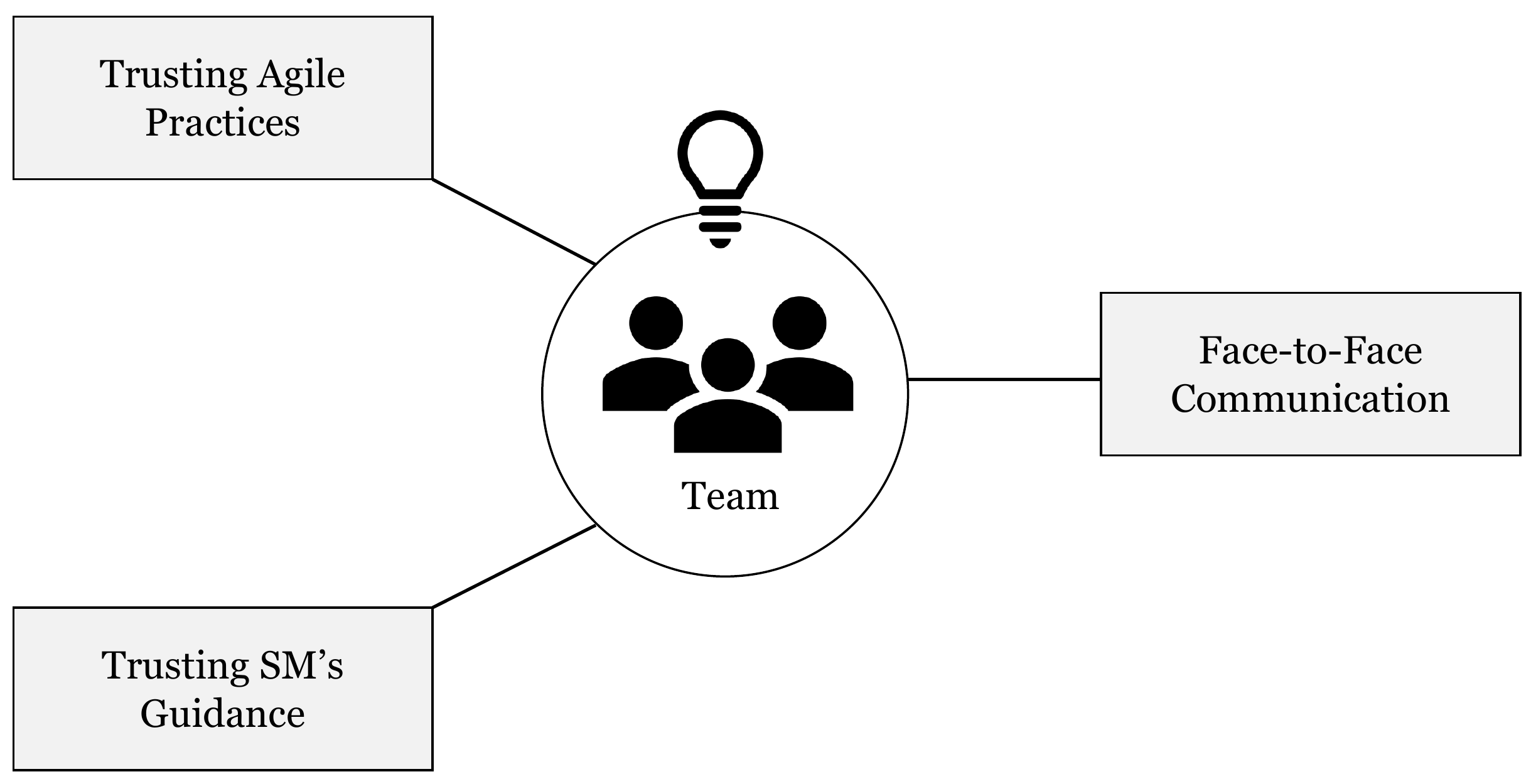}
        \caption{Strategies Used by the Team}
        \label{fig:teamStrategies}
        \end{minipage}
    \end{figure}

    \paragraph{\textbf{Having a Dedicated SM}}
    Having two/rotating SMs is challenging as described in the previous section. 
    When a team member plays the role of a developer and SM simultaneously, it becomes harder for them to perform the responsibilities of both the roles. Having a dedicated SM benefits the SM and the team. The SM gains members trust and confidence over time performing their roles more effectively.
    \begin{center}
          \textit{``..only having a single Scrum Master as they have more time to understand the role better, and thus be able to carry out their job more effectively.''} -- Team Member (Team 6)
    \end{center}

    \paragraph{\textbf{Gaining Scrum Knowledge}}
   Students acknowledged the importance of practitioners guides~\citep{2015TheGuide, Deemer2012ThePrimer}. These resources provide  guidelines for the students. Even though knowledge is given to the students through the course, students have reported that familiarity with Scrum guides on hand facilitates the SM. However, theoretical knowledge is not enough in playing the role effectively. Therefore, experience will sharpen and enhance the Scrum knowledge, but going through practitioner guides while playing the role allow the SM to play the role effectively and also to understand the concepts better.
    \begin{center}
       \textit{ ``I would suggest the Scrum Master to use practitioner guides as a general guideline on performing their duties and implementing the artefacts.''} -- Team Member (Team 6)
    \end{center}
    
    \paragraph{\textbf{Training the SM}}
SM being a university student, did not have enough experience and knowledge. The course content covers many aspects, introducing students to the role and responsibilities of the SM. Given the course's time constraint, it is not feasible to fully train the students to play each role. One of the objectives behind allowing the students to collaborate with the industry through this project is to get hands-on experience and training. This experience and knowledge will help them become better SM when they step into the industry after graduation. A trained SM plays a significant role in project success.
    \begin{center}
       \textit{ ``Inexperience can be solved with some training, but because the university course does not provide that much time, so the students often might not have the time to participate in all the training required. ''} -- Team Member (Team 7)
    \end{center}
    
    \paragraph{\textbf{Having a Dedicated Location for Meetings}}
 One of the responsibilities of SM is to allocate a dedicated location for meetings that is accessible to everyone in the team. Attendance in these meetings is essential. It is the SM who ensures that the team members share the project updates in these meetings. Time-boxing these meetings is an effective practice. Remote communication introduces several challenges both for the SM and team and to avoid them SM organises a physical meeting.
    \begin{center}
      \textit{  ``To get everyone on the same page, the Scrum Master could organise physical meetings where the team members can work together.'' } -- Team Member (Team 7)
    \end{center}
    
    \subsubsection{Most Common Strategies Used by the Team (RQ3b)}
    The strategies used by the teams to overcome the challenges in association with SM role are trusting agile practices, trusting SM's guidance, and having face-to-face communication as given in the Figure \ref{fig:teamStrategies}.
    
    \paragraph{\textbf{Trusting Agile Practices}}
    Our analysis implies that it is required for the team to trust agile practices to support SM's role. A team may decide to change the practices but as students reflected, it is important to maintain practising and trusting the agile practices. Having alternative techniques can be used to adhere with the practices to help the SM to play the role effectively. The flexible nature of agile allows the teams to change accordingly to fit to the particular situations, but also by letting the teams practice agile naturally. One such incident as reported by students is having less frequent daily stand-ups and using virtual communication channels. In a university context, this may be quite suitable as it is difficult for the students as they are not in a working environment during all working days. However, having the trust in agile practices but moulding accordingly help the teams to be benefited from agile.
    \begin{center}
      \textit{  ``To get the most benefit from Scrum, it is important for you to not be discouraged in modifying the Scrum practices to fit your particular team's workflow. For example, Daily stand-up is supposed to take place every day in person, however since that is not always possible and effective, I would recommend reducing the frequency and sometimes occasionally taking the virtual alternative, such as Messenger or Slack.''} -- Team Member (Team 4)
    \end{center}
    
    \paragraph{\textbf{Trusting SM's Guidance}}
    In order to have a better environment and a better relationship with the SM, it is necessary for the team to trust SM's guidance and allow the SM to play the role freely. Even though the students found that it is challenging for the teams to give opportunities for the SM to play the role accurately, but it is one of the ways that they can get the maximum benefits of SM's role.
    \begin{center}
       \textit{ ``I believe the challenges for the team would be to do our best to give opportunities for the Scrum Master to do their role.''} -- Team Member (Team 3)
    \end{center}
    
    \paragraph{\textbf{Having Face-to-Face Communication}}
    Communication matters the most in agile. Even though the teams have practised remote communication, students stated that face-to-face communication as a team is required. The reason for this as they found is that remote communication depletes the value of having Scrum events such as stand-ups, sprint planning, reviews, and retrospective meetings. The students also had confirmed this through literature. Therefore, the SM needs to facilitate events in a way that face-to-face communication is maintained.
    \begin{center}
       \textit{ ``Face-to-face communication cannot be beaten by any other method when having meetings. Many companies in the literature, and from my personal experience, have noted that voice or video communication is objectively inferior for standups, sprint planning, reviews, or retrospective meetings.''} -- Team Member (Team 2)
    \end{center}

    \subsection{Scrum Master Strategy Vs. Effectiveness of the Role}
    By considering all the responses to the 5 reflection surveys (N=331), we found that the students have rated the effectiveness of the SM role in each sprint within the range of 4-5 for all the SM strategies played in each sprint, \textit{one dedicated SM, two dedicated SM,} and \textit{rotating SM} in highest as given in the Table \ref{tab:strategyVsEffectivess}. However, even though the strategy \textit{one dedicated SM}'s highest response belongs to the range 4-5, it is only 69.3\%. Therefore, we performed regression analysis as well to come to a conclusion on the relationship between the SM strategy and the effectiveness of the role. The regression analysis confirms that there is no correlation between the strategy and the effectiveness, as the correlation coefficient (Multiple R=0.222215115) is closer to 0. This value should be near to 1 in order to have a strong positive relationship, and should be near to -1 for the relationship to be strongly negative.

\begin{table}[t]
    \resizebox{0.9\textwidth}{!}{%
    \begin{tabular}{@{}lcccc@{}}
    \toprule
   {Scrum Master Strategy} & \multicolumn{3}{l}{Role Effectiveness Rating Range} & \multicolumn{1}{l}{{Total Responses}} \\ \cmidrule(lr){2-4}
     & 1-2 & 3 & 4-5 & \multicolumn{1}{l}{} \\ \midrule
    One Dedicated Scrum Master & 9.1\% & 21.6\% & \textbf{69.3\%} & 88 \\
    Two Dedicated Scrum Masters & 1.5\% & 29.2\% & \textbf{87.7\%} & 178 \\
    Rotating Scrum Master & 2.8\% & 10.7\% & \textbf{82.6\%} & 65 \\ \bottomrule
    \end{tabular}%
    }
    \caption{Scrum Master Strategy Vs. Effectiveness of the Role (N=331), 1-2=low effectiveness, 3=average effectiveness, 4-5=high effectiveness}
    \label{tab:strategyVsEffectivess}
    \end{table}

    \subsection{Relationships among Challenges and Strategies}
\begin{figure}[t]
            \centering
            \includegraphics[width=\textwidth]{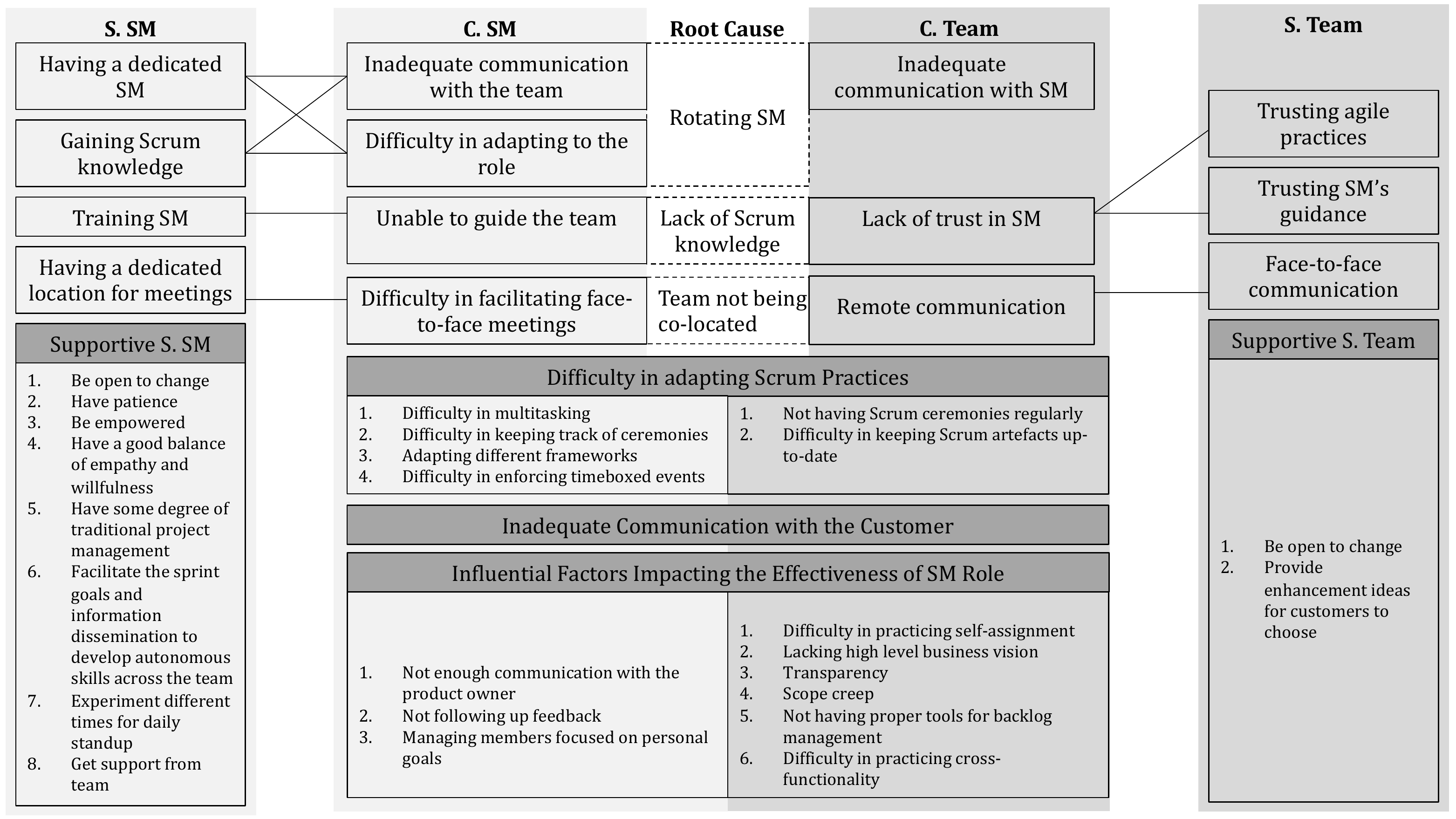}
            \caption{Relationships among Challenges and Strategies (S: Strategy; C: Challenge; SM: Scrum Master)}
            \label{fig:relationship}
        \end{figure}

    We identified that some root causes connect the challenges and strategies faced by both SM and the team. And also, to smoothen the performance of the SM's role, SM and team practice various strategies. Figure \ref{fig:relationship} provides an overview of the common root causes, challenges and strategies, and other strategies and challenges associated with the SM role.
    
    \subsubsection{Common Root Causes}
As shown in Figure \ref{fig:relationship}, certain challenges faced by the SM and team share the same root cause. Due to the presence of a rotating SM, both SM and team were challenged by having inadequate communication with each other as a transitioning period was required from one SM to another and team to get used to the new SM. Students pointed out that having a dedicated SM would solve this issue. Also, due to the same root cause, it has become difficult for the SM to get adapted to the role. Therefore, students have identified gaining Scrum knowledge, and training SM as the strategies to mitigate this challenge.

As the SM was lacking Scrum knowledge, he/ she was not able to the guide the team. This has led the team to doubt their trust on the SM. Gaining Scrum knowledge, and training the SM would build SM's confidence; thereby allowing the SM to guide the team smoothly. Trusting agile practices, and trusting SM's guidance were mentioned by the students as the strategies from team's side to overcome this issue.

Remote communication has been a challenge to the team members. As the team members were not co-located, it has become difficult for the SM to facilitate face-to-face meetings; which has steered to conduct meetings remotely.  The only solution to get control of this challenge as stated by the students was face-to-face communication. Therefore, having a dedicated, easy to access location was identified as a potential solution from SM's side by the students.

\subsubsection{Common Challenges}
Both SM and team found it difficult to adapt Scrum practices in early weeks due to lack of experience. It was challenging for the SM to keep track of ceremonies, adapt to different frameworks such as Scrumban, and to enforce timeboxed events. Conversely, teams struggled to have regular Scrum ceremonies, which was a consequence of SM having a difficulty to keep track of the ceremonies. Also, teams faced challenges in keeping the Scrum artefacts up-to-date. Both SM and team have been challenged by not having enough experience, and not having enough communication with the customer. However, strategies to overcome these challenges were not stated by the students.

\subsubsection{Influential Factors Impacting the Effectiveness of the Scrum Master Role}
Our analysis triggered that there are factors which influence the effectiveness of the SM role other than the challenges SM and team face. They found factors are inadequate communication with the product owner, not following up feedback on delivered product from the customer, and managing members focused on personal goals. Furthermore, from the perspective of the team, difficulty in practising self-assignment, lacking high level business vision, transparency, scope creep, not having proper tools for backlog management, and difficulty in practising cross-functionality were identified as factors impacting the effectiveness of the SM role. Yet, the counter-factors to minimise the mentioned factors were not identified by the students.

\subsubsection{Supportive Strategies}
Several supportive strategies by both SM and team were mentioned by the students to reinforce the role of SM. Both parties being open to change is one such strategy. Students have found that, having patience, being empowered, and having a good balance of empathy and wilfulness which we identify as interpersonal skills as strategies that SMs practice. Also, having some degree of traditional project management, facilitating the sprint goals and information dissemination to develop autonomous skills across the team, experimenting different times for daily stand-ups, and getting support from the team improve the quality of the SM's role. Additionally, students have mentioned that providing different ideas to customers to enhance the software that they develop, as a strategy that is being practised by the team.

\section{Overall Student Experience in the Course}
\label{Exp}

        \begin{figure}[t]
        \centering
        \includegraphics[scale=0.5]{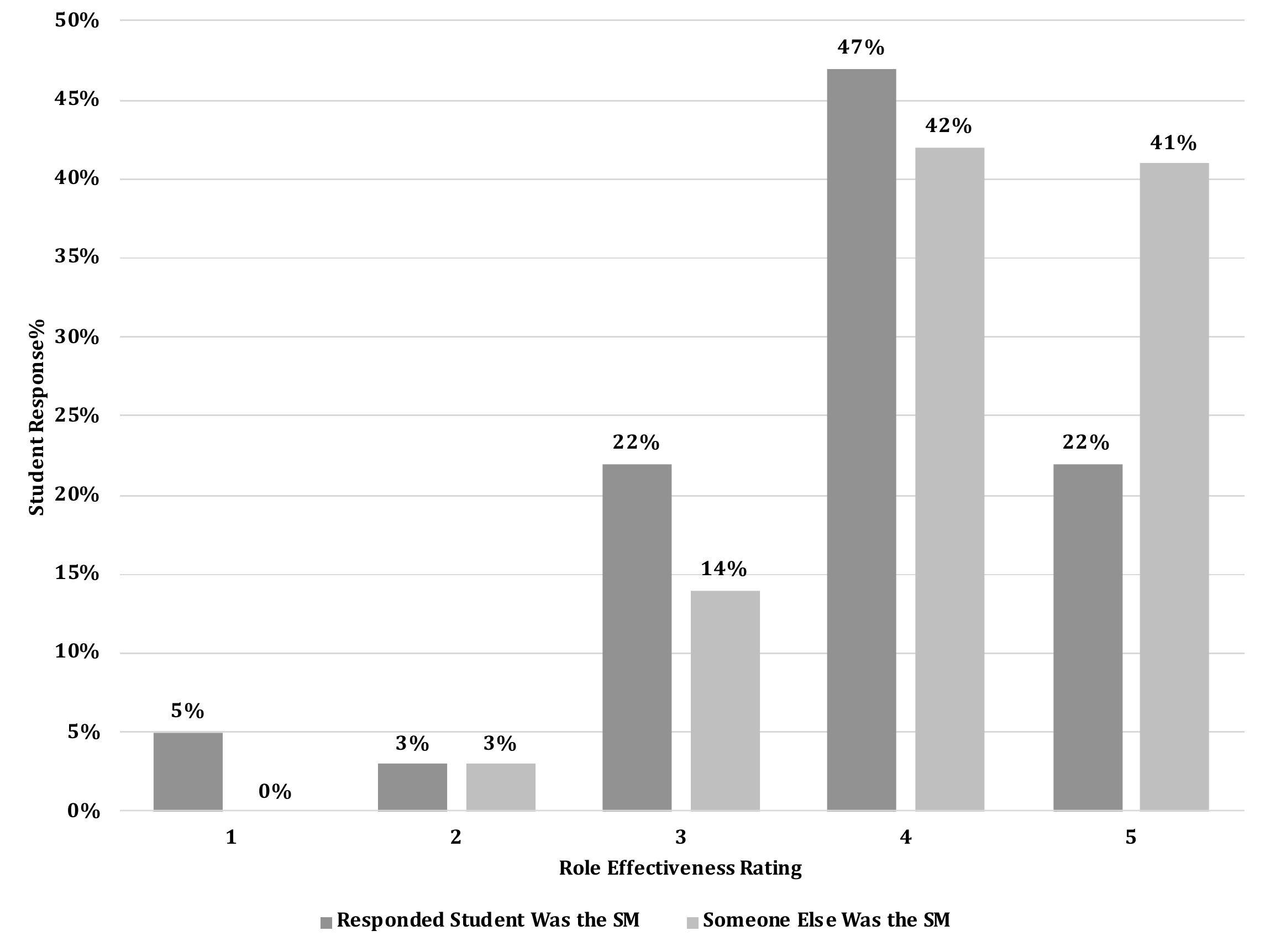}
        \caption{Scrum Master Strategy Vs. Effectiveness of the Role (SM: Scrum Master; N=331)}
        \label{fig:playedSM}
    \end{figure}
       
    \subsection{Student Perception on Playing the Role by Themselves Vs. Effectiveness of the Role}
    In order to discover how students measured the effectiveness of the role when they played it by themselves and when played by some other member in the team, we analysed data of all weeks (N=331). As shown in Figure \ref{fig:playedSM}, the majority of students rated themselves as 4 out of 5 when they played the role of SM. Also, the rate of effectiveness was reported highest 4 -- 5 out of 5 when played by someone else. From the graph, it is clear that students have not underestimated themselves when rating the effectiveness. But it is unclear whether they have overestimated themselves as well.
    
    \subsection{Student Satisfaction with the Customer Vs. Project Outcomes}
\begin{figure}[]
            \centering
            \includegraphics[scale=0.5]{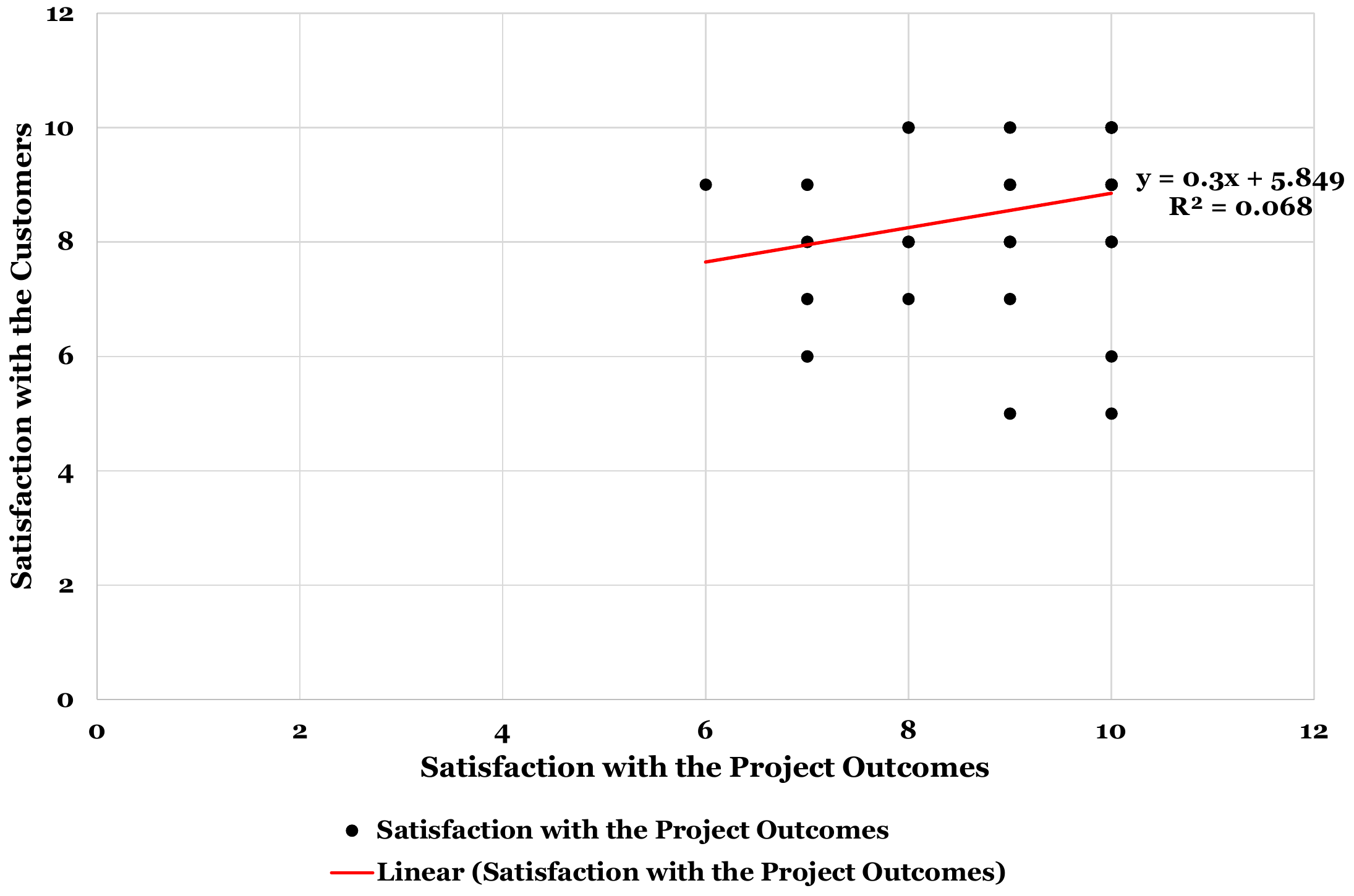}
            \caption{Correlation Between Student Satisfaction With the Customer and the Project Outcomes (N=51)}
            \label{fig:satisfacation}
        \end{figure}

    Regression analysis was carried out to distinguish whether a dependency exists on student satisfaction with the customer and the project outcomes. Similar to the other correlation findings, we found that there is no relationship between student's satisfaction with the customers and project outcomes as the correlation coefficient (Multiple R=0.260796) is ~0. This is further illustrated in the Figure \ref{fig:satisfacation}. The points deviate in an uneven manner from the trend line, visualising the clear nonexistence of the relationship between student satisfaction with the customer and project outcomes.
    
    \subsection{Contribution to Improvement in Agile Knowledge by the Course}
    We conducted an additional survey at the end of the course to have an understanding on students' improvement in agile knowledge through the course. As given in Table \ref{tab:contributionKnowledge} considering the highest rates, the students having medium agile knowledge have reported equally that their agile knowledge was ``somewhat'' and ``significantly'' improved (46\%). Majority of the students (53\%) who have claimed that they are with high agile knowledge, reported that their knowledge was ``somewhat'' improved by the course. Most students (57\%) with ``very high'' knowledge stated that their agile knowledge was improved ``significantly'' by the course. As the majority of the students were satisfied by the course, it can be concluded that delivery of this course where theory and practice play side-by-side was mostly successful.
    
        \begin{table}[t]
        \resizebox{\textwidth}{!}{%
        \begin{tabular}{@{}lcclcc@{}}
        \toprule
        {Prior Agile Knowledge} & \multicolumn{4}{l}{Contribution to Improvement in Agile Knowledge by the Course} & \multicolumn{1}{l}{{Total Responses}} \\ \cmidrule(lr){2-5}
         & \multicolumn{1}{l}{Not at all} & \multicolumn{1}{l}{No, Hardly} & Yes, Somewhat & \multicolumn{1}{l}{Yes, Significantly} & \multicolumn{1}{l}{} \\ \midrule
        Medium & 8\% & 0\% & \textbf{46\%} & \textbf{46\%} & 13 \\
        High & 0\% & 0\% & \textbf{53\%} & 47\% & 32 \\
        Very High & 0\% & 14\% & 29\% & \textbf{57\%} & 7 \\ \bottomrule
        \end{tabular}%
        }
        \caption{Contribution to Improvement in Agile Knowledge by the Course}
        \label{tab:contributionKnowledge}
        \end{table}

\section{Recommendations for Educators}
\label{Rcmd}
Educators are recommended to provide additional reading resources on the role of the SM to the students. They should encourage the industry collaborators to guide the students from their practical experience. They need to remind the industry collaborators to reduce the development load on the students who are playing the SM role. Teams are suggested to have two dedicated student SMs for a longer duration to share the responsibilities and gain experience. We recommend students with agile experience to take up the role of SM at the beginning of the project as it will set an example for the other students. Having tutors with SM expertise, knowledge, and skills will enhance students' experience with the role of the SM. Moreover, as students reported that working on other courses hindered their agile project work, educators are recommended to keep into account students commitments and their courses deadlines to help them plan and deliver the project work.

\section{Threats to Validity and Future Work}
This study was conducted with the participation of software engineering students in New Zealand. Therefore, the findings are particularly applicable to New Zealand but not generalisable to the global community. In the future, we encourage researchers to replicate this study in other geographical locations. This will allow the educators to compare and contrast the findings and build collective knowledge on the role of the SM in higher education contexts.

Due to the systems week (a final year requirement), the flow of the course conduct was affected. This may have affected the consistency of students’ involvement in the industry collaborated projects. Therefore, in the future, as similar courses are designed, we encourage the educators to minimise such pattern-breakers in the middle of the course or consider adapting their agile practices to suit such university specific constraints \citep{Masood2018AdaptingContexts}.

This study was conducted before the Covid--19 pandemic, an era where not having face--to--face meetings and remote communication were seen as challenges. However, with the pandemic, work from home and virtual meetings became commonplace. These came with their own set of challenges such as reduced productivity, communication and coordination challenges \citep{Miller2021} The long term benefits and challenges of virtual meetings as compared to co-located meetings remain to be seen.

\section{Conclusion}
Through this industry collaborated university course, we identified responsibilities of the Scrum Master, challenges faced by the SM and team along with the strategies to overcome the challenges. As reported by the students, the SM faces the challenges of: inadequate communication with the team, difficulty in adapting to the role, not being able to guide the team, difficulty in facilitating face-to-face meetings, difficulty in multitasking, difficulty in adapting Scrum practices, and inadequate communication with the customer. Sharing the latter three challenges mentioned above, teams face the challenges of inadequate communication with the SM, lack of trust in the SM, and remote communication.

The identified strategies to overcome these challenges are having a dedicated SM, gaining Scrum knowledge, training the SM, and having a dedicated location for the meetings, from the perspective of the SM. As a team, the students have stated that the team has to trust agile practices and guidance, and have face-to-face communication as much as possible. Our analysis reported supportive strategies such as being open to change, having patience, and be empowered, and more. In addition to that, the influential factors impacting the effectiveness of the SM role are also given in this paper. Additionally, we found that certain challenges faced by the SM and team share common root causes.

Given the challenges and strategies around the role of the SM identified by the students and their overall experience, we recommend educators to consider the recommendations we have provided to enhance student experience in agile projects.

\section*{Acknowledgement}
We thank all the SoftEng761 students at The University of Auckland, New Zealand, for sharing their experiences gained throughout the course.

\section*{CRediT Author Statement}
\textbf{Kashumi Madampe:} Formal Analysis, Data Curation, Writing - Original Draft, \& Editing, Visualisation, Project administration. \textbf{Zainab Masood:} Formal Analysis, Writing - Original Draft, Review \& Editing. \textbf{Rashina Hoda:} Conceptualization, Methodology, Investigation, Writing - Review \& Editing, Supervision.

\bibliographystyle{elsarticle-harv} 
\bibliography{main}

\begin{thebibliography}{29}
\expandafter\ifx\csname natexlab\endcsname\relax\def\natexlab#1{#1}\fi
\providecommand{\url}[1]{\texttt{#1}}
\providecommand{\href}[2]{#2}
\providecommand{\path}[1]{#1}
\providecommand{\DOIprefix}{doi:}
\providecommand{\ArXivprefix}{arXiv:}
\providecommand{\URLprefix}{URL: }
\providecommand{\Pubmedprefix}{pmid:}
\providecommand{\doi}[1]{\href{http://dx.doi.org/#1}{\path{#1}}}
\providecommand{\Pubmed}[1]{\href{pmid:#1}{\path{#1}}}
\providecommand{\bibinfo}[2]{#2}
\ifx\xfnm\relax \def\xfnm[#1]{\unskip,\space#1}\fi
\bibitem[{Bass(2014)}]{Bass2014ScrumProjects}
\bibinfo{author}{Bass, J.M.}, \bibinfo{year}{2014}.
\newblock \bibinfo{title}{{Scrum master activities: Process tailoring in large
  enterprise projects}}, in: \bibinfo{booktitle}{Proceedings - 2014 IEEE 9th
  International Conference on Global Software Engineering, ICGSE 2014}.
\newblock \DOIprefix\doi{10.1109/ICGSE.2014.24}.
\bibitem[{Bolloju et~al.(2018)Bolloju, Chawla and
  Ranjan}]{Bolloju2018ProsStudy}
\bibinfo{author}{Bolloju, N.}, \bibinfo{author}{Chawla, R.},
  \bibinfo{author}{Ranjan, R.}, \bibinfo{year}{2018}.
\newblock \bibinfo{title}{{Pros and cons of rotating scrum master role - A
  qualitative study}}, in: \bibinfo{booktitle}{ACM International Conference
  Proceeding Series}.
\newblock \DOIprefix\doi{10.1145/3172871.3172883}.
\bibitem[{Bunse et~al.(2004)Bunse, Feldmann and
  D{\"{o}}rr}]{Bunse2004AgileEducation}
\bibinfo{author}{Bunse, C.}, \bibinfo{author}{Feldmann, R.L.},
  \bibinfo{author}{D{\"{o}}rr, J.}, \bibinfo{year}{2004}.
\newblock \bibinfo{title}{{Agile methods in software engineering education}},
  in: \bibinfo{booktitle}{Lecture Notes in Computer Science (including
  subseries Lecture Notes in Artificial Intelligence and Lecture Notes in
  Bioinformatics)}.
\bibitem[{Cowan(2011)}]{Cowan2011WhenRunning}
\bibinfo{author}{Cowan, C.L.}, \bibinfo{year}{2011}.
\newblock \bibinfo{title}{{When the VP is a scrum master, you hit the ground
  running}}, in: \bibinfo{booktitle}{Proceedings - 2011 Agile Conference, Agile
  2011}.
\newblock \DOIprefix\doi{10.1109/AGILE.2011.31}.
\bibitem[{Davidson and Klemme(2016)}]{Davidson2016WhyMaster}
\bibinfo{author}{Davidson, A.}, \bibinfo{author}{Klemme, L.},
  \bibinfo{year}{2016}.
\newblock \bibinfo{title}{{Why a ceo should think like a scrum master}}.
\newblock \DOIprefix\doi{10.1108/SL-11-2015-0086}.
\bibitem[{Deemer et~al.(2012)Deemer, Benefield, Larman and
  Vodde}]{Deemer2012ThePrimer}
\bibinfo{author}{Deemer, P.}, \bibinfo{author}{Benefield, G.},
  \bibinfo{author}{Larman, C.}, \bibinfo{author}{Vodde, B.},
  \bibinfo{year}{2012}.
\newblock \bibinfo{title}{{The Scrum Primer}}.
\newblock \bibinfo{journal}{InfoQ} .
\bibitem[{Devadiga(2017)}]{Devadiga2017SoftwareIndustry}
\bibinfo{author}{Devadiga, N.M.}, \bibinfo{year}{2017}.
\newblock \bibinfo{title}{{Software Engineering Education: Converging with the
  Startup Industry}}, in: \bibinfo{booktitle}{Proceedings - 30th IEEE
  Conference on Software Engineering Education and Training, CSEE and T 2017}.
\newblock \DOIprefix\doi{10.1109/CSEET.2017.38}.
\bibitem[{Deved{\v{z}}i{\'{c}} and
  Milenkovi{\'{c}}(2011)}]{Devedzic2011TeachingStudy}
\bibinfo{author}{Deved{\v{z}}i{\'{c}}, V.}, \bibinfo{author}{Milenkovi{\'{c}},
  S.R.}, \bibinfo{year}{2011}.
\newblock \bibinfo{title}{{Teaching agile software development: A case study}}.
\newblock \bibinfo{journal}{IEEE Transactions on Education}
  \DOIprefix\doi{10.1109/TE.2010.2052104}.
\bibitem[{Digital.ai(2020)}]{202014thAgile}
\bibinfo{author}{Digital.ai}, \bibinfo{year}{2020}.
\newblock \bibinfo{title}{{14th State of Agile Report | State of Agile}}.
\newblock \bibinfo{type}{Technical Report}.
\bibitem[{Germain and Robillard(2005)}]{Germain2005Engineering-basedStudy}
\bibinfo{author}{Germain, Ã.}, \bibinfo{author}{Robillard, P.N.},
  \bibinfo{year}{2005}.
\newblock \bibinfo{title}{{Engineering-based processes and agile methodologies
  for software development: A comparative case study}}.
\newblock \bibinfo{journal}{Journal of Systems and Software}
  \bibinfo{volume}{75}, \bibinfo{pages}{17--27}.
\newblock \DOIprefix\doi{10.1016/j.jss.2004.02.022}.
\bibitem[{Hans(2017)}]{Hans2017WorkProjects}
\bibinfo{author}{Hans, R.T.}, \bibinfo{year}{2017}.
\newblock \bibinfo{title}{{Work in Progress - The Impact of the Student Scrum
  Master on Quality and Delivery Time on Students' Projects}}, in:
  \bibinfo{booktitle}{Proceedings - 5th International Conference on Learning
  and Teaching in Computing and Engineering, LaTiCE 2017}.
\newblock \DOIprefix\doi{10.1109/LaTiCE.2017.22}.
\bibitem[{Heikkil{\"{a}} et~al.(2016)Heikkil{\"{a}}, Paasivaara and
  Lassenius}]{Heikkila2016TeachingGame}
\bibinfo{author}{Heikkil{\"{a}}, V.T.}, \bibinfo{author}{Paasivaara, M.},
  \bibinfo{author}{Lassenius, C.}, \bibinfo{year}{2016}.
\newblock \bibinfo{title}{{Teaching university students Kanban with a
  collaborative board game}}, in: \bibinfo{booktitle}{Proceedings -
  International Conference on Software Engineering}.
\newblock \DOIprefix\doi{10.1145/2889160.2889201}.
\bibitem[{Hoda(2019)}]{hoda2019using}
\bibinfo{author}{Hoda, R.}, \bibinfo{year}{2019}.
\newblock \bibinfo{title}{Using agile games to invigorate agile and lean
  software development learning in classrooms}, in: \bibinfo{booktitle}{Agile
  and Lean Concepts for Teaching and Learning}. \bibinfo{publisher}{Springer},
  pp. \bibinfo{pages}{391--414}.
\bibitem[{Hoda(2021)}]{Hoda2021Socio-TechnicalEngineeringb}
\bibinfo{author}{Hoda, R.}, \bibinfo{year}{2021}.
\newblock \bibinfo{title}{Socio-technical grounded theory for software
  engineering}.
\newblock \bibinfo{journal}{IEEE Transactions on Software Engineering} ,
  \bibinfo{pages}{1--1}\DOIprefix\doi{10.1109/TSE.2021.3106280}.
\bibitem[{Kropp et~al.(2014)Kropp, Meier, Mateescu and
  Zahn}]{Kropp2014TeachingCollaboration}
\bibinfo{author}{Kropp, M.}, \bibinfo{author}{Meier, A.},
  \bibinfo{author}{Mateescu, M.}, \bibinfo{author}{Zahn, C.},
  \bibinfo{year}{2014}.
\newblock \bibinfo{title}{{Teaching and learning agile collaboration}}, in:
  \bibinfo{booktitle}{2014 IEEE 27th Conference on Software Engineering
  Education and Training, CSEE and T 2014 - Proceedings}.
\newblock \DOIprefix\doi{10.1109/CSEET.2014.6816791}.
\bibitem[{Masood et~al.(2018)Masood, Hoda and
  Blincoe}]{Masood2018AdaptingContexts}
\bibinfo{author}{Masood, Z.}, \bibinfo{author}{Hoda, R.},
  \bibinfo{author}{Blincoe, K.}, \bibinfo{year}{2018}.
\newblock \bibinfo{title}{{Adapting agile practices in university contexts}}.
\newblock \bibinfo{journal}{Journal of Systems and Software}
  \bibinfo{volume}{144}, \bibinfo{pages}{501--510}.
\newblock \DOIprefix\doi{10.1016/j.jss.2018.07.011}.
\bibitem[{Matturro et~al.(2015)Matturro, Font{\'{a}}n and
  Raschetti}]{Matturro2015SoftMasters}
\bibinfo{author}{Matturro, G.}, \bibinfo{author}{Font{\'{a}}n, C.},
  \bibinfo{author}{Raschetti, F.}, \bibinfo{year}{2015}.
\newblock \bibinfo{title}{{Soft skills in scrum teams: A survey of the most
  valued to have by product owners and scrum masters}}, in:
  \bibinfo{booktitle}{Proceedings of the International Conference on Software
  Engineering and Knowledge Engineering, SEKE}.
\newblock \DOIprefix\doi{10.18293/SEKE2015-026}.
\bibitem[{Melnik and Maurer(2003)}]{Melnik2003IntroducingLearned}
\bibinfo{author}{Melnik, G.}, \bibinfo{author}{Maurer, F.},
  \bibinfo{year}{2003}.
\newblock \bibinfo{title}{{Introducing agile methods in learning environments:
  Lessons learned}}.
\newblock \bibinfo{journal}{Lecture Notes in Computer Science (including
  subseries Lecture Notes in Artificial Intelligence and Lecture Notes in
  Bioinformatics)} .
\bibitem[{Miller et~al.(2021)Miller, Rodeghero, Storey, Ford and
  Zimmermann}]{Miller2021}
\bibinfo{author}{Miller, C.}, \bibinfo{author}{Rodeghero, P.},
  \bibinfo{author}{Storey, M.A.}, \bibinfo{author}{Ford, D.},
  \bibinfo{author}{Zimmermann, T.}, \bibinfo{year}{2021}.
\newblock \bibinfo{title}{{"How Was Your Weekend?" Software Development
  Teams Working From Home During COVID-19}}.
\newblock \DOIprefix\doi{10.1109/ICSE43902.2021.00064}.
\bibitem[{Noll et~al.(2017)Noll, Razzak, Bass and Beecham}]{Noll2017ARole}
\bibinfo{author}{Noll, J.}, \bibinfo{author}{Razzak, M.A.},
  \bibinfo{author}{Bass, J.M.}, \bibinfo{author}{Beecham, S.},
  \bibinfo{year}{2017}.
\newblock \bibinfo{title}{{A study of the scrum master's role}}, in:
  \bibinfo{booktitle}{Lecture Notes in Computer Science (including subseries
  Lecture Notes in Artificial Intelligence and Lecture Notes in
  Bioinformatics)}.
\newblock \DOIprefix\doi{10.1007/978-3-319-69926-4{\_}22}.
\bibitem[{Paasivaara(2021)}]{paasivaara2021teaching}
\bibinfo{author}{Paasivaara, M.}, \bibinfo{year}{2021}.
\newblock \bibinfo{title}{Teaching the scrum master role using professional
  agile coaches and communities of practice}, in: \bibinfo{booktitle}{2021
  IEEE/ACM 43rd International Conference on Software Engineering: Software
  Engineering Education and Training (ICSE-SEET)},
  \bibinfo{organization}{IEEE}. pp. \bibinfo{pages}{30--39}.
\bibitem[{Paasivaara et~al.(2015)Paasivaara, Blincoe, Lassenius, Damian,
  Sheoran, Harrison, Chhabra, Yussuf and Isotalo}]{Paasivaara2015LearningTeams}
\bibinfo{author}{Paasivaara, M.}, \bibinfo{author}{Blincoe, K.},
  \bibinfo{author}{Lassenius, C.}, \bibinfo{author}{Damian, D.},
  \bibinfo{author}{Sheoran, J.}, \bibinfo{author}{Harrison, F.},
  \bibinfo{author}{Chhabra, P.}, \bibinfo{author}{Yussuf, A.},
  \bibinfo{author}{Isotalo, V.}, \bibinfo{year}{2015}.
\newblock \bibinfo{title}{{Learning Global Agile Software Engineering Using
  Same-Site and Cross-Site Teams}}, in: \bibinfo{booktitle}{Proceedings -
  International Conference on Software Engineering}.
\newblock \DOIprefix\doi{10.1109/ICSE.2015.157}.
\bibitem[{Paasivaara et~al.(2013)Paasivaara, Lassenius, Damian, Raty and
  Schroter}]{Paasivaara2013TeachingScrum}
\bibinfo{author}{Paasivaara, M.}, \bibinfo{author}{Lassenius, C.},
  \bibinfo{author}{Damian, D.}, \bibinfo{author}{Raty, P.},
  \bibinfo{author}{Schroter, A.}, \bibinfo{year}{2013}.
\newblock \bibinfo{title}{{Teaching students global software engineering skills
  using distributed Scrum}}, in: \bibinfo{booktitle}{Proceedings -
  International Conference on Software Engineering}.
\newblock \DOIprefix\doi{10.1109/ICSE.2013.6606664}.
\bibitem[{Rico and Sayani(2009)}]{Rico2009UseEducation}
\bibinfo{author}{Rico, D.F.}, \bibinfo{author}{Sayani, H.H.},
  \bibinfo{year}{2009}.
\newblock \bibinfo{title}{{Use of agile methods in software engineering
  education}}, in: \bibinfo{booktitle}{Proceedings - 2009 Agile Conference,
  AGILE 2009}.
\newblock \DOIprefix\doi{10.1109/AGILE.2009.13}.
\bibitem[{Schwaber and Sutherland(2015)}]{2015TheGuide}
\bibinfo{author}{Schwaber, K.}, \bibinfo{author}{Sutherland, J.},
  \bibinfo{year}{2015}.
\newblock \bibinfo{title}{{The Scrum Guide}}, in: \bibinfo{booktitle}{Software
  in 30 Days}.
\newblock \DOIprefix\doi{10.1002/9781119203278.app2}.
\bibitem[{Shastri et~al.(2021)Shastri, Hoda and Amor}]{yogi2020scrummaster}
\bibinfo{author}{Shastri, Y.}, \bibinfo{author}{Hoda, R.},
  \bibinfo{author}{Amor, R.}, \bibinfo{year}{2021}.
\newblock \bibinfo{title}{Spearheading agile: The role of the scrum master in
  agile projects}.
\newblock \bibinfo{journal}{Empirical Software Engineering}
  \bibinfo{volume}{26}, \bibinfo{pages}{1--31}.
\bibitem[{Shukla and Williams(2002)}]{Shukla2002AdaptingCourse}
\bibinfo{author}{Shukla, A.}, \bibinfo{author}{Williams, L.},
  \bibinfo{year}{2002}.
\newblock \bibinfo{title}{{Adapting extreme programming for a core software
  engineering course}}, in: \bibinfo{booktitle}{Software Engineering Education
  Conference, Proceedings}.
\newblock \DOIprefix\doi{10.1109/CSEE.2002.995210}.
\bibitem[{Waugh(2018)}]{Waugh2018IncreasingCertification}
\bibinfo{author}{Waugh, M.}, \bibinfo{year}{2018}.
\newblock \bibinfo{title}{{Increasing effectiveness of library projects through
  scrummaster certification}}, in: \bibinfo{booktitle}{Advances in Library
  Administration and Organization}.
\newblock \DOIprefix\doi{10.1108/S0732-067120180000038018}.
\bibitem[{Yi(2011)}]{Yi2011ManagerMaster}
\bibinfo{author}{Yi, L.}, \bibinfo{year}{2011}.
\newblock \bibinfo{title}{{Manager as scrum master}}, in:
  \bibinfo{booktitle}{Proceedings - 2011 Agile Conference, Agile 2011}.
\newblock \DOIprefix\doi{10.1109/AGILE.2011.8}.

\end{thebibliography}

\end{document}